\documentclass[11pt]{article}

\sffamily
\usepackage{pifont}
\newcommand{\cmark}{\ding{51}}%
\newcommand{\xmark}{\ding{55}}%
\usepackage{booktabs}
\usepackage{authblk}
\usepackage{hyperref} 
\usepackage{cite}
\usepackage{epsfig}
\usepackage{bm}
\usepackage{amssymb}
\usepackage{mathrsfs}
\usepackage{amsmath}
\usepackage{dsfont}
\usepackage{color}

\newcommand{\be}{\begin{equation}}
\newcommand{\ee}{\end{equation}}
\newcommand{\cl}{\notag\\}
\newcommand{\Q}{Q}

\newcommand{\bx}{x}

\newcommand{\lat}{\text{lat}}

\usepackage{cleveref}

\begin{document}

\thinmuskip=3mu					
\medmuskip=4mu 					
\thickmuskip=4mu plus 3mu minus 1mu  	

\title{\Large{
%
Abelian gauge theories on the lattice: $\theta$-terms and compact gauge theory with(out) monopoles
}\vskip10mm}

\vspace{35mm}

\author[1]{Tin Sulejmanpasic\thanks{\href{mailto:tin.sulejmanpasic@gmail.com}{tin.sulejmanpasic@gmail.com}}}

\author[2]{Christof Gattringer\thanks{\href{mailto:christof.gattringer@uni-graz.at}{christof.gattringer@uni-graz.at}}}

\affil[1]{\small Philippe Meyer Institute, Physics Department, \'Ecole Normale Sup\'erieure,\newline 
PSL Research University, 24 rue Lhomond, F-75231 Paris Cedex 05, France \vspace{3mm} }

\affil[2]{\small University of Graz, Institute for Physics,  A-8010 Graz, Austria \vspace{2mm}}

\begin{titlepage}
\maketitle
\thispagestyle{empty}

\begin{abstract}

We discuss a particular lattice discretization of abelian gauge theories in arbitrary dimensions. 
The construction is based on gauging the center symmetry of a non-compact abelian gauge theory, 
which results in a Villain type action. We show that this construction has several benefits over the 
conventional $U(1)$ lattice gauge theory construction, such as electric-magnetic duality, natural 
coupling of the theory to magnetically charged matter in four dimensions, complete control over the 
monopoles and their charges in three dimensions and a natural $\theta$-term in two dimensions. 
Moreover we show that for bosonic matter our formulation can be mapped to a worldline/worldsheet 
representation where the complex action problem is solved. We illustrate our construction by explicit 
dualizations of the $CP(N\!-\!1)$ and the gauge Higgs model in $2d$ with a $\theta$ term, as well as 
the gauge Higgs model in $3d$ with constrained monopole charges. These models are of importance 
in low dimensional anti-ferromagnets. Further we perform a natural construction of the $\theta$-term 
in four dimensional gauge theories, and demonstrate the Witten effect which endows magnetic matter 
with a fractional electric charge. We extend this discussion to $PSU(N)=SU(N)/\mathbb Z_N$ non-abelian 
gauge theories and the construction of discrete $\theta$-terms on a cubic lattice. 

\end{abstract}

\end{titlepage}

\tableofcontents
\newpage
\section{Introduction}

An important aspect of many quantum field theories is the fact that they can be augmented with topological 
$\theta$-terms which can cause dramatic changes in 
the low-energy physics. The corresponding physical phenomena are non-perturbative in nature such that 
genuinely non-perturbative approaches are necessary to study these systems. 

Such terms are well known in the context of non-abelian gauge theories, as they are the source of the famous 
strong $CP$ problem. However $\theta$-terms are also a feature of many condensed matter systems, often 
in the context of effective abelian gauge theories. A famous example is the spin-chain described by an 
effective $U(1)$ abelian gauge theory with a $\theta$-term. The $\theta$-terms also appear in the context 
of topological insulators (see, e.g., \cite{Qi:2008ew,Essin:2008rq,Seiberg:2016rsg}).

In higher dimensions $U(1)$ gauge theories generically have monopoles. In fact in the conventional lattice 
formulations of $U(1)$ gauge theories monopoles seem like an unavoidable curse of the formalism. 
Yet while many effective theories have monopoles, they do not appear as unavoidable, and their presence 
and/or charge is dictated by the underlying symmetry of the UV theory. For the example of (2+1)$d$ 
anti-ferromagnets, a good effective theory is a $U(1)$ gauge Higgs theory, but with monopoles of a 
particular charge (see \cite{haldane1988,Baskaran1988, Murthy:1989ps,senthil2004a,senthil2004b,Sachdev:2015slk}). 
The constraint is imposed by the presence of $\mathbb Z_N$ lattice symmetries, which are mapped 
into the conservation of the magnetic flux modulo $2\pi N$ in the corresponding $(2+1)d$ effective quantum field theory. 
This symmetry is sometimes referred to as the \emph{magnetic center symmetry} or the topological symmetry\footnote{The terms topological is sometimes used 
because the symmetry is a remnant of the topological $U(1)$ symmetry for which the current $j=\frac{1}{2\pi}\star F$ is 
conserved identically by the Bianchi identity, not due to the Noether theorem. However the distinction is not a physical one, and this symmetry is a usual Noether symmetry in other dual representations of the theory.  }.. This requirement means 
that the effective theory cannot be allowed to have monopoles (i.e., space-time flux changing events) whose 
charge is not a multiple of $N$.

All of the scenarios above have interesting setups where partial answers about their long-distance properties 
are provided by anomaly matching 
\cite{Gaiotto:2017yup,Komargodski:2017smk,Tanizaki:2017qhf,Sulejmanpasic:2018upi,Tanizaki:2018xto,Hongo:2018rpy}. 
In this work we are interested in formulating such theories on the lattice in a way amenable to numerical 
simulations. To achieve this, the discretization on the lattice has to be done in a way such that the 
symmetries of the continuum theories are implemented exactly in order to compare to the anomaly 
matching constraints. Standard lattice discretization often does not meet these criteria. For example the 
$U(1)$ gauge theory in $(1+1)d$ can be supplemented by a $\theta$-term, but a crucial anomaly matching 
ingredient is the preservation of charge-conjugation symmetry at $\theta=\pi$ 
\cite{Gaiotto:2017yup,Sulejmanpasic:2018upi,Tanizaki:2018xto}. In $(2+1)d$ there are also anomaly matching 
conditions involving a discrete subgroup of the $U(1)$ magnetic center (or topological) symmetry 
\cite{Komargodski:2017dmc,Komargodski:2017smk,Tanizaki:2018xto}, the preservation of which requires monopole 
charges to be multiples of some integer other than unity. The conventional lattice discretization is also 
difficult to couple to monopole matter in a satisfactory way. It is also not clear that it enjoys an electric-magnetic 
duality, well known for abelian gauge theories in the continuum.

The second obstacle towards a successful lattice treatment is the fact that appropriate definition of the 
path-integral representation may still suffer from the complex action problem that has to be overcome 
before the formulation can be used in a Monte Carlo simulation. Consequently the lattice discretization 
of the theory has to be such that it admits a solution of the complex action problem if the theory is to be useful for simulations. This can sometimes be achieved by a transformation 
to dual variables which are worldlines for matter fields and worldsheets for the gauge degrees of freedom.

In this paper we revisit the problem of finding suitable lattice discretizations in order to properly define and 
simulate the abelian gauge theories discussed above. Our construction is based on a formulation of a 
non-compact abelian gauge theory, and subsequent gauging of its $\mathbb R$ center symmetry down to 
$U(1)$. The procedure results, up to a gauge fixing, in a Villain action for the gauge field. As we will see, this 
approach will allow us to couple magnetic matter, and implement an electric-magnetic duality even in the 
presence of electric and magnetic matter. It also serves as a natural way to define $\theta$-terms in both 
2$d$ and 4$d$ lattice gauge theories.

We will also discuss the application of this construction to three model systems: 
The $2d$ gauge Higgs model with a $\theta$-term, the 2d $CP(N\!-\!1)$ 
model with a $\theta$-term, and the $3d$ gauge Higgs model with a new type of term that allows to control the 
charge of the monopoles. For all three model systems we work out the details of the lattice formulation and 
discuss the corresponding dual worldline/worldsheet representation where the complex action problem is 
resolved\footnote{We remark that in some parts of our discussion, in particular Sections 2 and 3, we 
use the language of differential forms on the lattice, and we summarize our corresponding 
notations in Appendix A.}.

The paper is organized as follows. Sec.~\ref{compact_noncompact} is devoted to the general discussion 
about the discretization we use in all dimensions, and the motivation behind it. Here we discuss electric and magnetic matter-matter coupling, and dualization. In $4d$ the theory has self dual points with interesting consequences. In  Sec.~\ref{sectionGHtheta} we discuss the $CP(N-1)$ nonlinear sigma model 
and the $U(1)$ gauge Higgs model in $2d$. In Sec.~\ref{sectionGHmonopoles} we focus on $3d$ 
theories with constrained magnetic charges. In Sec.~\ref{sec:4d} we discuss $\theta$-terms in $4d$, and 
the Witten effect associated with such terms.

\section{Gauging center symmetry and lattice discretization} 
\label{compact_noncompact}

\subsection{Compact vs.\ non-compact abelian gauge theory}

To motivate the steps we later take for the lattice regularization, we begin with a review of 
compact vs.\ non-compact formulations of abelian gauge theories in the continuum, 
as well as the connection between the two. We first consider non-compact abelian gauge 
theory, i.e., an $\mathbb R$-gauge theory on a $d$-dimensional torus $\mathbb T^d$. 
The minimal (Euclidean) gauge action is given by
\be
\frac{1}{4e^2}\int_{\mathbb T^d} \! d^{\, d}x \, F_{\mu\nu}^2 \; ,
\label{SG_continuum}
\ee
where the field strength is given by 
$ F_{\mu\nu}=\partial_\mu A_\nu-\partial_\nu A_\mu$ and $A_\mu$ is the gauge field. 
Making use of the differential form language\footnote{Here we use the notation of differential forms, where $F=\frac{1}{2!}F_{\mu\nu} \;dx^\mu\wedge dx^\nu=\frac{1}{2!}F_{\mu\nu}\epsilon^{\mu\nu}d^2x$ in component form.} we may write
\be
F \; = \; dA \; ,
\ee
where the field strength $F$ is a 2-form and $A$ is the 1-form gauge field, with $d$ denoting the exterior derivative. 
We note that $F$ is a total derivative globally, meaning that $A$ is a globally well defined gauge field.

Due to the nilpotency of the exterior derivative ($d^2 = 0$), the theory is invariant under the gauge symmetry 
\be
A \; \rightarrow \; A \; + \; d \, \Omega \; .
\ee
It is crucial that $\Omega$ is a single valued function on the torus, i.e., the gauge transformation 
0-form $\Omega$ takes values in the non-compact group $\mathbb R$. This is in distinction to the 
compact formulation where the gauge transformation parameter $\Omega$ parametrizes a circle, which is only well defined $\bmod\; 2\pi$.

Note that the non-compact abelian gauge theory cannot have a global flux. 
Indeed, $A$ is a single valued gauge field on the torus, and the field strength is a total derivative, such that
\be
\int_{\mathbb T^d}\, F \, = \, 0\;.
\ee 

Wilson line operators are defined as
\be\label{eq:nc_wilson}
W_\gamma \; = \; e^{ik\oint_\gamma \, A}\; \; , \quad k\in \mathbb R \; ,
\ee
where $\gamma$ is some closed curve. Unlike the compact $U(1)$ gauge theory which we will discuss below, 
the non-compact gauge theory allows the charge $k$ of the test particle to take any real (non-quantized) value. 

Finally, there exists a symmetry transformation 
\be
A \; \rightarrow \; A \, + \, \alpha \; ,
\label{Ashift}
\ee
where $\alpha$ is required to obey $d\, \alpha=0$. 
In other words, $A$ can be shifted by a flat connection $\alpha$, without changing 
the action. The operators charged under this symmetry are Wilson loops along a non-contractable curve $\gamma$, 
and not local operators as is the case with the conventional global symmetries. As a consequence the conserved 
charge of such a symmetry is an integral over a co-dimension 2 hypersurface. The term \emph{1-form symmetry} 
was used in 
\cite{Kapustin:2014gua,Gaiotto:2014kfa,Gaiotto:2017yup} to distinguish such symmetries from the usual (0-form) 
global symmetries, which have local operators charged under them. We will also refer to 
(\ref{Ashift}) as the \emph{electric center symmetry}, because the objects charged under it are electrically charged Wilson loops.

Let us now switch to compact $U(1)$ gauge theory. The action is the same as before, except that now 
gauge transformations are given by
\be
A\, \rightarrow \; A \, + \, d \, \omega \; ,
\ee
where $\omega$ is only defined modulo $2\pi$ (i.e., it is the parameter of the gauge group $U(1)$). 
Clearly, now the gauge invariant Wilson loops are well defined only for integer charges,
\be
W_\gamma \; = \; e^{ik\oint_\gamma \, A}\; \; , \quad k \in \mathbb Z\;,
\ee
i.e., test charges $k$ are quantized in integer units. Also it can easily be seen that the electric center symmetry is 
now a $U(1)$ symmetry. 

Furthermore, $F$ is now a closed, but not an exact 2-form (if the theory does not contain dynamical monopoles), and the description in terms of a global gauge field 
$A$ is not admissible. In particular the theory allows for a non-trivial flux through a $2$-cycle $\Gamma$,
\be\label{eq:magnetic_flux}
\int_{\Gamma}  \, F \; \in \; 2\pi \, \mathbb Z \;.
\ee
As a result the operator
\be\label{eq:mag_sym_gen}
e^{\frac{\theta}{2\pi} i\int_\Gamma  \, F}
\ee
is a non-trivial operator in the compact gauge theory. This operator is a topological 
$\theta$-term in $d$ = 2 space-time dimensions. 

In higher dimensions, the operator \eqref{eq:mag_sym_gen} it is a generator of the \emph{magnetic center symmetry}. The operators charged under this symmetry are monopole operators which are co-dimension 3 operators (i.e., 
spacetime points in 3$d$, lines in 4$d$, sheets in 5$d$, etc.), which we label as $M_q(\mathcal C)$, where 
$\mathcal C$ is a 
co-dimension 3 hypersurface. The defining property of the monopole operator $M_q(\mathcal C)$ is that
\be
e^{i\frac{\theta}{2\pi} \int_{\Gamma} \, F} \, M_q(\mathcal C)\; = \; e^{i\,\theta q \, \ell(\Gamma,\mathcal C)}
\; M(\mathcal C) \; ,
\ee
where $\Gamma$ is a surface, and $\ell(\Gamma, \mathcal C)$ is the linking number of $\Gamma$ and $\mathcal C$. 
In other words $\ell (\Gamma,\mathcal C)=1$ if the surface $\Gamma$ wraps once around the monopole operator 
along $\mathcal C$. Another way of saying this is that we require that an arbitrary surface wrapping around a 
monopole has $q$-units of flux \eqref{eq:magnetic_flux}.

If monopoles are not dynamical, and we have a pure $U(1)$ gauge theory, then the operator \eqref{eq:magnetic_flux} 
is clearly a conserved quantity. This can also be seen from the Bianchi identity
\be
dF \; = \; 0 \;,
\ee
so that the 2-form current
\be
j \; = \; \frac{1}{2\pi} \; \star F
\ee
is conserved, where $\star$ is the Hodge star operator. The objects charged under this symmetry are the monopole 
operators, which are ($d-3$)-dimensional operators, so the symmetry is a ($d-3$)-form symmetry (see the 
discussion in \cite{Gaiotto:2014kfa}).

Now we introduce matter fields, or more specifically electric matter\footnote{We will introduce magnetic matter 
in the lattice gauge theory version, but will not discuss it in the continuum.}. We will discuss only the bosonic case, 
but a similar discussion holds for fermionic matter. In both cases we introduce matter fields via
terms in the action of the form 
\be
\sum_{i} |D_\mu(k_i) \phi_i|^2 \; + \; \dots \; ,
\ee
where $D_\mu(k_i)=\partial_\mu+ik_iA_\mu$ is the covariant derivative with charge $k_i$ and  the dots indicate 
 gauge-invariant interaction terms of $\phi_i$. Clearly the fields $\phi_i$ transform as charge-$k_i$ matter fields under the gauge 
transformation. If $A$ is a non-compact gauge field then the charges $k_i$ can be arbitrary real numbers. However, in 
the case of the compact gauge field, the charges can only be integer\footnote{We always normalize our gauge field so that the flux $F$ is normalized in $2\pi\mathbb Z$. }. This is because a gauge transformation $A
\rightarrow A+d\omega, \phi_i\rightarrow e^{-i k_i\omega} \phi_i$ is inconsistent with the single-valuednes of 
$\phi_i(x)$ on a torus if $k_i$ is not integer.

The presence of matter reduces the center symmetry. In particular the center symmetry acts as 
$A\rightarrow A+\alpha$, where $d\alpha=0$. We can always write $\alpha = d\rho$, where $\rho$ is not single valued on the manifold, so that the transformation may affect the Wilson loops. This means that 
$\rho$ is not necessarily well defined globally. If we can now transform $\phi_i\rightarrow e^{-i \rho k_i}\phi_i$ 
in a way such that $e^{-i \rho k_i}$ is well defined globally for all $k_i$, then we can say that there exists a 
residual center symmetry. Indeed, let us assume that all ratios of charges are rational numbers 
$k_i/k_j\in \mathbb Q, \forall i,j$. Then there exists a constant $c\in \mathbb R$ such that $k_i/c\in \mathbb Z$. 
Now define $\rho=\omega/c$, where $\omega(x)$ is well defined modulo $2\pi$, i.e., it is an angle. Then 
\be
e^{-i \rho k_i} \; = \; e^{-i\omega k_i/c}
\ee
is well defined, hence the center symmetry group is $\mathbb Z$. On the other hand, if there exist $i,j$, such 
that $k_i/k_j\in \mathbb R\backslash\mathbb Q$, the center symmetry is completely broken by the presence of matter fields.  

Similar reasoning can be used to see that choosing 
$\rho(x)=\frac{\omega(x)}{\text{GCD}(\{k_i\})}$ in the compact gauge field case, 
where $\omega(x)$ is well defined $\bmod\; 2\pi$, 
results in having a $Z_{\text{GCD}(\{k_i\})}$ center symmetry, where  
$\text{GCD}(\{k_i\})=\text{GCD}(k_1,k_2,\dots)$ denotes the greatest common divisor of integers $k_1,k_2,\dots \;$. 
We summarize the properties of compact and non-compact gauge theories in Table~\ref{tab:noncomp_vs_comp}.
\begin{table}[t!]
   \centering
   \begin{tabular}{@{} lcr @{}} 
      \toprule
    \midrule
     				& Non-compact & Compact\\
        \midrule
      Flux $\int F$      & $0$ & $2\pi\mathbb Z$\\
      Allowed electric charge k     & $k\in \mathbb R$  & $k\in \mathbb Z$ \\
      Monopole operators   &  \xmark & \cmark \\
      Center sym. (pure gauge) & $\mathbb R$ & $U(1)$\\
      Center sym. (with $\{k_i\}$ charge matter fields) & $\mathbb Z$ or $\mathbb Z_1$ & 
      $\mathbb Z_{\text{GCD}(\{k_i\})}$\\
      \bottomrule
   \end{tabular}
   \caption{Summary of differences between compact and non-compact gauge theory.}
   \label{tab:noncomp_vs_comp}
\end{table}

In fact one can obtain the $U(1)$ compact gauge theory from a non-compact gauge theory by gauging a discrete 
center symmetry. Because this is best understood in the context of lattice field theory, we will discuss this procedure 
upon lattice discretization. In fact it is this procedure which will allow us to formulate a lattice discretization which is 
useful for our goals.

\subsection{The lattice discretization: Gauge fields}

For our discretization we consider a cubic lattice $\Lambda$ with periodic boundary conditions 
and denote the lattice sites as $x$, lattice links as $l$, 
plaquettes as $p$, cubes as $c$, hypercubes $h$ and general $r$-cells as $c^{(r)}$, such that $x \sim c^{(0)}$, $l \sim c^{(1)}$, 
$p \sim c^{(2)}$, $c \sim c^{(3)}$  and $h\sim c^{(4)}$. In parts of this paper we will borrow the language of discrete differential
forms and the theory of simplicial complexes adapted to cubic lattices. Furthermore, some of our notation is similar, 
but not identical to previous work (see, e.g., \cite{Ukawa:1979yv,Yaffe:1979iq,Kapustin:2014gua,Kapustin:2013qsa}). 
We also emphasize that our construction is for cubic lattices, rather than simplicial complexes (i.e., triangulations of
manifolds) like in \cite{Kapustin:2013qsa}. In Appendix \ref{app:forms} we provide a summary of the details of our notation 
for discrete differential forms on cubic lattices.

We define $U(1)$ non-compact gauge theory coupled to scalar fields on the lattice using the gauge degrees of 
freedom $A_l\in \mathbb R$ living on the links $l$ of the lattice and matter fields $\phi_x \in \mathbb C$ living 
on the lattice sites. A natural gauge-kinetic action of this theory is given by
\be\label{eq:noncompact_action}
S \; = \; \frac{\beta}{2} \, \sum_p F_p^{\, 2} \; ,
\ee
where the sum is over the plaquettes $p$ of a given orientation, which we take as positive\footnote{If no other indication is given, it is assumed that the orientations of links $l$, plaquettes $p$, cubes $c$ and general $r$-cells $c^{(r)}$ are the same and positive.}.  $F_p$ is a discretization of the field 
strength tensor, which may be chosen as the oriented sum along the links in the boundary of $p$, 
\be
F_p \; = \; \sum_{l\in \partial p} \, A_l \; .
\label{Fp}
\ee
$\partial p$ is the set of all links that bound the plaquette $p$, taking into account the orientation of the links (see Appendix \ref{app:forms}). 
We also identify $l$ and $- l$ as the same link but with opposite orientation, and we have 
chosen a convention in which $A_{-l}=-A_{l}$. If we denote a link $l$ by its base site $x$ and its direction 
$\mu$, and a plaquette $p$ with its base point $x$ and its two directions $\mu$ and $\nu$, the above definition amounts to 
\be
F_{x,\mu\nu} \; = \; A_{x+\mu,\nu} \, - \, A_{x,\nu} \, - \, A_{x+\nu,\mu} \, + \,A_{x,\mu} \; ,
\ee
which is a natural discretization of the field strength tensor. As we summarize in Appendix A,
Eq.~(\ref{Fp}) defines an exterior derivative operator on link fields, i.e.,
\be
F_p \; = \; (dA)_p \; .
\ee
An analogous exterior derivative operator $d$ can be defined for any ``higher-form'' fields living 
on $r$-cells of the lattice, as we show in Appendix A. 

The action has a gauge symmetry, given by 
\be
A_l \; \rightarrow \; A_l \, + \, (d\Omega)_l \; ,
\ee
where $\Omega_x$ is an arbitrary lattice field on sites, and gauge invariance follows from the identity $d^2=0$ 
(see Appendix A). 

\begin{figure}[t] 
   \centering
   \includegraphics[width=4in]{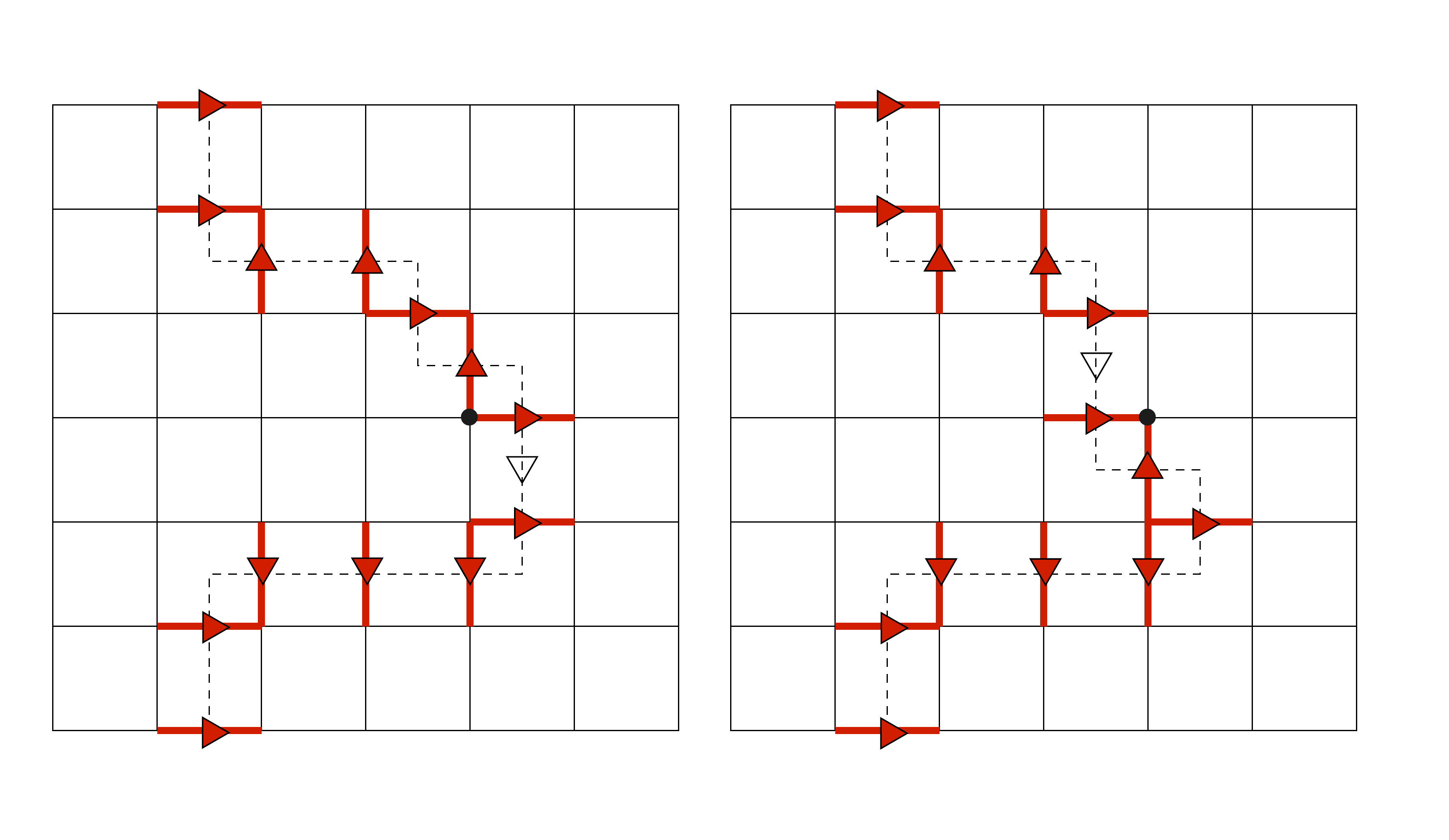} 
   \caption{Two examples of gauge-equivalent center-symmetry transformations. The dashed line is the 
   co-dimension 1 hyper-surface $\mathcal S$ along the links of the dual lattice. The links where the gauge 
   field is shifted by a constant are marked in red. Notice that the orientation of the links is fixed so that the intersection number between  $\mathcal S$ and the links is positive. The figure on the left and right are related by a 
   gauge transformation which has its support on the site labeled by a black dot.}
   \label{fig:center_sym}
\end{figure}

Gauge invariant Wilson loops are given by
\be
W_\gamma \; = \; \prod_{l\in \mathcal \gamma }e^{ik \, A_l} \; ,\; k\in \mathbb R \; ,
\ee
where $\gamma$ is some closed oriented contour built out of lattice links. 

Let us now pick an oriented co-dimension 1 hyper-surface $\mathcal S$ on the dual lattice 
(an example in $2d$ is given by the dashed line in Fig.~\ref{fig:center_sym}). 
Let $\mathcal L_{\mathcal S}$ be the set of all links which intersect $\mathcal S$, such that the 
intersection number between every link in  $\mathcal L_{\mathcal S}$ and $\mathcal S$  is positive. Now consider the following  transformation
\begin{subequations}
\begin{align}
&A_l \; \rightarrow \; A_l \, + \, \alpha, \;\alpha\in\mathbb R\;,\; \forall  l \in \mathcal L_{\mathcal S} \; ,\\
&A_l \; \rightarrow A_l, \;  \forall l\notin \mathcal L_{\mathcal S} \; .
\end{align}
\end{subequations}
Obviously, this transformation leaves the action invariant, but may affect Wilson loops. More specifically,
if the contour $\gamma$ has a non-vanishing sum $N_{\gamma,\mathcal{S}}$ of oriented 
intersections with $\mathcal S$ (i.e., each intersection is counted with $\pm 1$ depending on whether intersection of $\gamma$ and $\mathcal S$ is right-hand or left-hand oriented), then the Wilson loop transforms as:
\begin{equation}
W_\gamma \; \rightarrow \; e^{\, i \alpha k \, N_{\gamma,\mathcal{S}}} \;
W_\gamma \; .
\end{equation}
Note, however, that we can always perform a gauge transformation to deform the hyper-surface $\mathcal S$: 
For example if $x_0$ is the site marked with a black dot in Fig.~\ref{fig:center_sym}, we can perform 
a gauge transformation defined by $\Omega_{x_0} = -\alpha$ and 
$\Omega_x = 0 \; \forall x \neq x_0$ and transform the dashed contour in the lhs.\ plot of 
Fig.~\ref{fig:center_sym} into the one on the rhs. Thus it is clear that the center symmetry is a gauge 
tranformation if $\mathcal S$ is contractable. If $\mathcal S$ is not contractible, then the transformation 
corresponds to a global symmetry transformation which we call an \emph{electric center symmetry transformation}. 

So far our center symmetry transformation was an $\mathbb R$ symmetry\footnote{More accurately for the 
$d$-dimensional torus it is a direct product of $d$ copies of $\mathbb R$, one for each 1-cycle of the torus. In general 
the center symmetry group of the $\mathbb R$ field is the 1st cohomology group of the manifold with coefficients 
in $\mathbb R$, i.e., $H^1(\mathcal M,\mathbb R)$. This is made obvious either by noting that $\delta$ is 
a representative of $H^1(\mathcal M,\mathbb R)$, or that the hyper-surface $\mathcal S$ is a member of the 
$(d\!-\!1)$-st homology of $H_{d-1}(\mathcal M,\mathbb R)$, which by Poincare duality is isomorphic to 
$H^1(\mathcal M,\mathbb R)$.}. Now we wish to construct a theory with a $U(1)$ center symmetry. 
We do this by gauging the subgroup $\mathbb Z$ of the center symmetry $\mathbb R$. In other words we now 
want the transformations which correspond to shifts of $A_l$ by $2\pi \, \mathbb Z$ to be gauge transformations.  
Choosing the center transformation as $\alpha_l \, = \, 2\pi \, n_l$ with $n_l \in \mathbb{Z}$ we find that 
the field strength transforms as
\begin{subequations}\label{eq:1-form_gauge_transform}
\be
F_p \; \rightarrow \; F_p \, + \, 2 \pi \, (d\,n)_p \; ,
\ee
which is clearly not an invariance of the action \eqref{eq:noncompact_action}. To make this an invariance of 
the action we introduce a discrete gauge field $B_p \in 2\pi \, \mathbb Z$ based on plaquettes $p$, and promote 
the field strength to 
\begin{equation}
F_{p} \; \rightarrow \; F_p \, + \, B_p \; .
\end{equation}
Now we demand that under the transformation $A_l\rightarrow A_l+2\pi \, n_l$ our discrete gauge field
$B_p$ transforms as
\be\label{Bp_gaugetrafo}
B_p \; \rightarrow \; B_p \; - \; 2\pi \, (dn)_p\; .
\ee
\end{subequations}
Thus when using the gauged field strength $F_p \, + \, B_p$ in the action, this action 
is gauge invariant under the transformation $A_l\rightarrow A_l+2\pi n_l$. 

In fact we have obtained  a compact gauge theory, with action
\be\label{eq:gauge_action_B}
S \; = \; \sum_{p}\frac{\beta}{2}(F_p+B_p)^2\;,
\ee
that, in addition to the link-based fields $A_l$, is endowed with a 2-form, $\mathbb Z$-valued gauge field 
$B_p$ based on the plaquettes.

In 2 dimensions we can deform the action by adding the gauge invariant term
\be\label{eq:action_theta}
S_\theta \; = \; i\frac{\theta}{2\pi} \, \sum_p B_p \; .
\ee
In fact this is precisely the $\theta$-term, as we will show in a moment. 

For more than 2 dimensions the field $B_p$ may have more degrees of freedom. 
In fact the introduction of the field $B_p$ on plaquettes now allows for
monopoles. We define a monopole operator as an operator on a cube $c$ (i.e., a $3$-cell) of the lattice. It is 
obvious that  
\be
(dB)_c \; = \; 2\pi \, q \;, \quad q\in \mathbb Z\; , 
\ee 
and for $q \neq 0$ we say that \emph{a monopole of charge $q$ sits on the cube $c$.} In $d=3$ dimensions 
this defines a monopole operator that is a point operator on the dual lattice.

In dimensions $d>3$ the monopole operator is an extended operator on the dual lattice consisting of adjacent 
cubes $c$ which span a $(d-3)$-dimensional hypersurface on the dual lattice (i.e., it is a particle in $4d$, 
a string in $5d$, et cetera).

Therefore, upon introducing the fields $B_p$  we have changed the theory in a significant way, as we now have 
locally very different configurations compared to what we had in the non-compact theory, and in $d>3$ there are 
even new states (particles, strings...) corresponding to the monopoles. All this is well known for the compact 
gauge theory. The conventional lore is that these monopoles are an inescapable 
curse of the abelian gauge theory on the lattice, but this is \emph{not} true in our formalism. Indeed we can 
choose to gauge the center symmetry with the field $B_p$ in such a way so that its curvature is flat, i.e., by 
imposing the constraint that
\be\label{eq:Bflux_constraint}
(dB)_c \; = \; 0 \; \; \; \forall c\;.
\ee
Now this means that $B$-flux is topologically conserved and quantized, which in turn implies the 
presence of a magnetic center (or topological) $U(1)$ symmetry. Indeed the object which 
transforms under that symmetry is precisely the monopole operator. 

Let us formulate this a bit more precisely. We impose the above constraint by introducing a delta 
function in the partition sum, which is equivalent to introducing the following term in the action:
\be
\label{eq:deformation}
i \, \Q \; = \; i \, \frac{1}{2\pi} \sum_{c} (\star  A^m)_{c}(dB)_{c}\;.
\ee
Here the sum is over all the cubes $c$ of the lattice, and $(\star  A^m)_c$ is a $\mathbb{R}$-valued 
3-form on the lattice $\Lambda$ 
which lives on the cubes $c$. We choose to think of this $3$-form as the $\star$-dual of the 
$(d-3)$-form  $ A^m$ on the dual lattice (see our definition of $\star$-duality in Appendix \ref{app:forms}). 
When integrating $(\star  A ^m)_c$ over the interval $[-\pi,\pi]$ at each cube, the Boltzmann factor with the 
exponent (\ref{eq:deformation}) gives rise to Kronecker deltas that impose the constraints \eqref{eq:Bflux_constraint} 
on all cubes $c$. The shift $(\star  A ^m)_c\rightarrow (\star  A^m)_{c}+\delta$ is a symmetry of the theory 
as long as $(d\delta)_{\tilde c^{(d-3)}}$. This can be seen by a use of partial integration formula from the 
Appendix \ref{app:forms}. Moreover, since $(\star  A^m)_{c}$ is an angular variable taking values 
$(\star  A^m)_{c}\in[-\pi,\pi)$, there is a $U(1)$ symmetry in the theory associated with the topological 
symmetry (see also a related discussion in 
\cite{Sulejmanpasic:2016uwq, Komargodski:2017dmc, Komargodski:2017smk}). 

The reason why it is useful to think of the Lagrange multiplier as the $\star$-dual of the $(d-3)$-form fields $ A^m$ 
on the dual lattice, is because $ A^m$ also enjoys an abelian gauge symmetry. Indeed if we transform
\be
 A^m \; \rightarrow \;  A^m \, + \, d \tilde \lambda \; ,
\ee
the action $Q$ remains unchanged,
\be
\Delta \Q \; = \; \frac{1}{2 \pi} \sum_{c} (\star d \tilde \lambda)_{c} (dB)_{c}  \; = \; 
-\frac{1}{2 \pi}  \sum_{h} \, (\star \tilde \lambda)_{h} (d^{\,2} B)_{h} \; = \; 0 \; ,
\ee
where we used the partial integration theorem\footnote{Notice that $\tilde\lambda=(-1)^{(d-4)(d+1)}\star^2\tilde\lambda$ 
(see Appendix \ref{app:forms} and the related discussion below).} (see  Appendix \ref{app:forms}).
We have therefore obtained another $U(1)$ $(d-3)$-form gauge field $ A^m$ which naturally lives on the dual 
lattice. This is a \emph{magnetic gauge field}. In $d=4$ the $U(1)$ gauge field is the usual 1-form gauge field,  and in $d=3$ it is just a scalar field on the dual lattice. 

To describe the corresponding monopole operator, we must take a product of $U(1)$ variables along a 
$(d-3)$-dimensional hyper-surface $\tilde{\mathcal S}$ built out of $(d-3)$-cells $\tilde c^{(d-3)}$ of the dual lattice, 
i.e.,
\be
T(\tilde{\mathcal S}) \, = \, \prod_{\tilde c^{(d-3)}\in \tilde{\mathcal S}}e^{i (\tilde A^m)_{\tilde c^{(d-3)}}}\;.
\ee
This is the 't Hooft loop operator of a monopole in $d$-dimensions.

\subsection{\texorpdfstring{$r$}{r}-form \texorpdfstring{$U(1)$}{U(1)} abelian gauge 
fields in \texorpdfstring{$d$}{d} dimensions}

We briefly comment on the construction of $U(1)$ $r$-form gauge fields in $d$ dimensions. Let us start with the 
$\mathbb R$-valued abelian $r$-form field $A_{c^{(r)}}$ where $r<d$, i.e., a real field that lives on the $r$-cells of 
$\Lambda$. In an analogous way as before we can define the gauge field kinetic term as
\be
\frac{\beta}{2} \sum_{c^{(r+1)}}  \big( (dA)_{c^{(r+1)}} \big)^2 \; .
\ee
The theory possesses an $r$-form $\mathbb R$-symmetry under the transformation  
$A_{c^{(r)}} \, \rightarrow \, A_{c^{(r)}} + \alpha_{c^{(r)}}$ where $\alpha_{c^{(r)}}$ is required to obey 
$(d \, \alpha)_{c^{(r+1)}}= 0$. As before we may lift this constraint and introduce an $(r+1)$-form $U(1)$ gauge field 
$B_{c^{(r+1)}} \in 2\pi\mathbb Z$. Requiring the transformation law 
$B_{c^{(r+1)}} \, \rightarrow \, B_{c^{(r+1)}} - (d \, \alpha)_{c^{(r+1)}}$, we find that the action 
\be
\sum_{c^{(r+1)}}\frac{\beta}{2}\big((dA)_{c^{(r+1)}} \, + \, B_{c^{(r+1)}}\big)^2
\ee
is invariant under the transformation $A_{c^{(r)}} \, \rightarrow \, A_{c^{(r)}} + \alpha_{c^{(r)}}$, where 
$\alpha_{c^{(r)}}$ is now an arbitrary $r$-form. 

In the special case of $r=0$, we have upon renaming
$A_{c^{(0)}} \Rightarrow \varphi_x$ and $B_{c^{(r+1)}} \Rightarrow 2 \pi n_l$, with $n_l \in \mathbb{Z}$, that
\be
\sum_{l}\frac{\beta}{2}\big((d\varphi)_l+2\pi n_l\big)^2\;,
\ee
which is just the Villain-type discretization of the $xy$ model in $d$-dimensions. 
By imposing the constraint $(dn)_p$ the model can be made vortex free. 


\subsection{The lattice discretization: Matter fields with electric and magnetic charge}
\label{discretization}

Now we discuss the lattice discretization of matter fields. Here we will consider only bosonic matter, although similar 
formulas can be derived for fermionic matter, with the caveat that some lattice formulations that do not suffer from the 
sign problem in bosonic formulation will suffer from it in the case of fermionic matter. 

We impose periodic boundary conditions for all fields on our lattice $\Lambda$ of size $N_s^{d-1} \times N_t$ 
(with the lattice spacing again set to 1).
The matter fields $\phi_x \in \mathbb{C}$ are assigned to the sites $x \in \Lambda$.
For coupling the gauge fields we introduce the $U(1)$-valued link variables 
\begin{equation}
U_{x,\mu} \; = \; e^{\, i \, A_{x,\mu}} \; ,
\label{Udef}
\end{equation}
where $A_{x,\mu}$ are the link-based gauge fields introduced before. Note that coupling the gauge fields via the
$U(1)$-valued link variables (\ref{Udef}) serves equally well for compact as well as for non-compact gauge fields.

The discretization for the matter field action is the conventional one: the derivatives in the continuum are 
replaced by nearest neighbor differences and we obtain the lattice action 
\begin{equation}
S_{e}^{\, \lat}[\phi,U] \; = \; \sum_{\bm x\in\Lambda}\left[ M_e |\phi^e_{x}|^2 + V_e(|\phi^e_{x}|) - \sum_{\mu=1}^{d} 
\Big({\phi^e_{{x}}}^{\ast} U_{{x},\mu} \phi^e_{{x}+{\mu}}+c.c.\Big) \right] \; ,
\label{S_Higgs}
\end{equation}
where the superscript/subscript $e$ stands for ``electric'' to emphasize that the matter carries an electric charge. We will sometimes drop this label for the electrically charged matter.
We defined the mass parameter $M_e$ as $M_e = 2d + m^2_e$,  where $m_e$ is the bare mass and the additional term
$2d$ comes from the on-site contributions in the discretized second derivatives. We will often switch between notations of labeling an $r$-cell by $c^{(r)}$ for $r$-cells, or by its base vertex $x$ and directions $\mu_1, \mu_2, \dots,\mu_r$ that span the $r$-cell. E.g., a link may either be denoted by $l$, by $c^{(1)}$ or by $(x,\mu)$, a plaquette by 
$p$, $c^{(2)}$ or  $(x,\mu \nu)$ et cetera (see Appendix A). 

In the $d=4$ case there is another kind of matter action that our gauge-field discretization allows. As we have shown, by insisting that the theory is monopole-free, we have introduced a dual-lattice, $(d-3)$-form gauge field $A^m$ living on the $c^{(d-3)}$-cells. These are just links of the dual lattice $\tilde\Lambda$ in $d=4$. So we can introduce lattice scalar 
(or fermion) fields which carry monopole charge in the same way as we did for the electric charge
\begin{equation}
{ S}_{m}^{\, \lat}[\phi^m,U^m] \; = \; \sum_{{\tilde{x}}\in\tilde \Lambda}\left[ M_m |\phi^m_{\tilde{ x}}|^2 + V_m(|\phi^m_{ {\tilde x}}|) - \sum_{\mu=1}^{d} 
\Big(\phi_{{{\tilde x}}}^{\ast} U^m_{{\tilde x},\mu} \phi^m_{{\tilde x}+{\mu}}+c.c.\Big) \right] \; ,
\label{S_Higgs_m}
\end{equation}
where we have introduced a dual-magnetic link variable $U_{{\tilde x},\mu}^m=e^{i A^m_{\tilde x,\mu}}$ 
in a way analogous to (\ref{Udef}).

The above magnetic-matter can only be formulated in $d=4$ dimensions, because only in $d=4$ the magnetic potential 
is an ordinary $1$-form gauge field which can be coupled naturally to matter fields. 

In $d=3$ the field $A^m$ is a compact scalar field on the dual lattice. We can therefore write a potential for $\sigma$ 
itself$\,$\footnote{In principle an action can be written for the $A^m$-field in any dimension, as long as it obeys gauge 
invariance. We will not bother writing these out here, but they may be useful for some applications.}, with the only 
restriction that it is periodic when $\sigma\rightarrow \sigma +2\pi$. So in general we have an action of the form
\be\label{eq:Monopole_Action}
 S^{\, \lat}_{m}[A^m]=-\sum_{n}\sum_{\tilde \bx}m_{n}\cos(nA^m_{\tilde\bx})\;.
\ee
If we choose $m_n=0$ whenever $n\ne 0\bmod\; N$ for some $N\in \mathbb Z$, then there is a remaining 0-form 
$\mathbb Z_N$ magnetic symmetry, sometimes also referred to as the topological symmetry.

In all dimensions we can define electric matter. The partition sum $Z_e^{\, \lat} [U]$ for the electric matter 
field in a background configuration $U$ of link variables is then given by
\begin{equation}
Z_e^{\, \lat} [U] \; = \; \int \! D[\phi] \; e^{-S^{\, \lat}_e[\phi,U]} \; ,
\label{ZM_lattice}
\end{equation}
where the path integral measure $\int \!  D[\phi]$ is defined as the product measure 
$\int \! D[\phi] = \prod_{x \in \Lambda} \int_\mathbb{C} d \phi_x / 2\pi$. In $d=4$ we 
analogously can define the partition sum $ Z_{m}^{\, \lat}$ for magnetic matter 
using $S_m^{\, \lat}$ instead of $S_e^{\, \lat}$.

We are now ready to finalize the construction and discretize the Boltzmann factor for the gauge field which also 
includes the term $\Q[A,A^m]$ defined in (\ref{eq:deformation}) which corresponds to the $\theta$-term for 
$d=2$ and to the magnetic-gauge-field coupling in higher dimensions. This Boltzmann factor is given by 
$e^{\, - S_G[A] \,- \, i \Q[B,A^m] }$ where both the gauge action $S_G[A]$ and the monopole term $\Q[B,A^m]$ we 
consider are functionals that have the form of an ``integral'' over 
space-time, i.e., discrete sums over sites, links and plaquettes. It is useful to define the Boltzmann factor
\begin{align}
B_G^{\,lat} [A,A^m]& =\prod_p \sum_{n_p\in\mathbb Z} e^{ - \sum_{p} \left[ \frac{\beta}{2} ( F_{p}  +  B_{p})^2  +  i \frac{1}{2\pi}B_{p}
{(\star{F}^m)}_{p} \right]} .
\label{B_villain}
\end{align}
The first exponent in (\ref{B_villain}) is the quadratic term from the gauge action, while the second term corresponds to \eqref{eq:deformation}. Here $F^m_{\tilde{c}_{d-2}}$ lives on $d-2$-cells of the dual lattice and is identified as the $(d-2)$-form gauge field strength of the $(d-3)$-form magnetic gauge field ${A}^m_{\tilde l}$ (see the discussion below Eq.~\eqref{eq:deformation}) in dimensions $d>2$, while for $d=2$ we can set $(\star F^m)_{\bm x,\mu\nu}=\theta \epsilon_{\mu\nu}$ to obtain the $\theta$-term (note however that the $A^m$ field is not defined in 2d) .

We will also often make more explicit the $2\pi$-periodicity of $B_P$ and write
\be
B_p=2\pi n_p, n_p\in \mathbb Z\;,
\ee
In the absence of the action $Q$ the 
Boltzmann factor reduces to the usual Villain form\footnote{There is a subtle difference between the Villain form and the form here. The way we obtained the kinetic action is by gauging the non-compact field theory, so that the link-fields take values $A_l\in(-\infty,\infty)$. However we can use the non-compact 1-form gauge symmetry to gauge fix the $A_l$ so that they take values in the interval $A_l\in(-\pi,\pi]$.} \cite{Villain:1974ir}, and for this reason we will sometimes refer to the plaquette-based 
integers $n_{p} \in \mathbb{Z}$ as ``Villain variables''.

The imaginary term in (\ref{B_villain}) has introduced a complex action problem, which in the next subsection we will solve
with an exact transformation to worldlines and worldsheets as the new dynamical degrees of freedom. Let us comment on this a bit. In $d>2$ in the absence of monopoles, the sign problem is easily solved by simply imposing the constraint $(dB)_{c}=0$. We can also impose a less stringent constraint that $(dB)_{c}=0\bmod\; q$, so that we only allow charge $q$-monopoles. But this would not control the dynamics of the monopoles very precisely, and would not account for their short-range interactions. In fact this too may be possible to account for by just adding terms in the action which depend on powers of $(dB)_{c}$. However, coupling the theory to magnetic matter allows an easy control over the monopole bare mass and the short-range interactions, as well as endowing them with fermionic or bosonic statistics by coupling the gauge field to either bosons or fermions. It also allows for formulations of theories which exhibit a nice electric-magnetic duality in four dimensions.

We complete our construction by defining the path integral measure for the non-compact gauge fields as the product
measure 
\begin{equation}
	\int \! \! D[A] \; = \; \prod_{x \in \Lambda} \, \prod_{\mu = 1}^d \,  \int_{-\infty}^{\infty} \frac{dA_{x,\mu}}{2\pi}  \;  ,
\label{Ameasure}	
\end{equation}
which for later use we have normalized with factors $1/(2\pi)$, as we will often gauge-fix to set $A_{x,\mu}\in[-\pi,\pi)$. The final form of the lattice discretization 
of the partition sum is then given by
\begin{equation}
Z^{\, \lat} \; = \; \int \!D[A^m]\int \!D[A] \; B_G^{\,lat} [A,A^m] \; Z_e^{\, \lat} [U] \tilde Z_{m}^{\, \lat}[U^m]\; ,
\label{Z_lat}
\end{equation}
with $B_G^{\,lat}$ defined in (\ref{B_villain}) and $Z_e^{\, \lat} [U]$ in (\ref{ZM_lattice}), with $\tilde Z_{m}^{\, \lat}[U^m]$ given for $d=4$ by a similar expression with the action of electric matter in \eqref{S_Higgs} replaced by the magnetic matter in \eqref{S_Higgs_m}. In $d=3$ we can define a magnetic monopole potential as in \eqref{eq:Monopole_Action}, and add it to the action. Each cosine term can then be expanded in the dual variables on the sites of the dual lattice (or cubes of the original lattice), and the dual-magnetic field $A^m$ can be integrated out, imposing constraints without introducing the sign problem as long as all $m_n$ in \eqref{eq:Monopole_Action} are positive. In the subsequent sections this general form of the lattice discretization with non-compact fields will be used and 
worked out in detail and applied to three examples.

\subsection{General form of the dual representation with worldlines and worldsheets}
\label{subsection_dual_general}

We now come to a general discussion of the dual representation in terms of worldlines and worldsheets 
that solves the complex action problem introduced by the term Eq.~\eqref{eq:deformation}. In subsequent sections
we discuss various models with our discretization. There
we will work out the corresponding specific dual form in more detail, but the general structure of the dualization 
is the same as the one here presented here.

The discussion here is for theories in $d$ dimensions which are described by the 
partition sum (\ref{Z_lat}) with the gauge Boltzmann factor (\ref{B_villain}) and some partition function 
$Z_e^{\, \lat}[U]$ for the matter sector, for example  the one defined by (\ref{ZM_lattice}) and (\ref{S_Higgs}). The details
of $Z^{\, \lat}_e[U]$ are actually irrelevant for the current discussion -- we only assume that it is a gauge invariant 
function of the link variables $U_{x,\mu}$ (for a more concrete example see, e.g.,
the discussion in \cite{Mercado:2013ola,Mercado:2013yta}, or also the short presentations
in  Appendices \ref{app:Higgs} and \ref{app:CPN}). 

The gauge invariant functions $Z_e^{\, \lat}[U]$ have a simple form: To maintain gauge invariance, the link
variables $U_{l}$ must form closed loops. 
The loops can be superimposed and a link 
can be visited several times. This gives rise to powers $(U_{l})^{k_{l}}$ of link variables, where the current variables (sometimes also referred to as the flux variables\footnote{The term flux may be slightly confusing here because, as we shall see, they arise because of the presence of an ordinary symmetry, which results in conservation of an ordinary current. The conservation of a 1-form symmetry will give rise to plaquette-valued integers which obey conservation of 2-form currents -- i.e., fluxes.}) $k_{l} \in \mathbb{Z}$ assigned to the links of the lattice denote how many
times a link $l=(\bx,\mu)$ is run through in either positive ($k_{l} > 0$) or negative ($k_{l} < 0$)  $\mu$-direction.  
A value $k_{l} = 0$ indicates an empty link $l$. 

The fact that the link variables must arrange along closed loops can be expressed as a zero divergence condition
for the flux variables $k_{x,\mu}$:
\begin{equation}
(\vec{\nabla} \cdot \vec{k})_x \; \equiv \; \sum_{\mu = 1}^d \big[ k_{x,\mu} - k_{x-\mu,\mu} \big] \; = \; 0 \; \; \; \; \forall x \; .
\end{equation}
The above condition can also be expressed more concisely as (see appendix)
\be
(\delta k)_x=(\star d\star k)_{x}=0\;,
\ee
i.e., $\star k$ is a flat (or closed) $(d-1)$-form  on the dual lattice.
The zero divergence condition ensures that the flux variables form closed loops and that the product 
of all link contributions $(U_{l})^{k_{l}}$ indeed gives a gauge invariant combination.

The matter partition sum $Z_e[U]$  then can be written in the form
\begin{align}\label{Ze_loop}
Z_e^{\, \lat} [U] 
& =  \sum_{\{k\}} \, W_e[k] \; \prod_x \delta \big( (\vec{\nabla} \cdot \vec{k})_x \big) \; 
\prod_{l} e^{\, i \, A_{l} \, k_{l}} \; ,
\end{align}
where $\delta(n)=\delta_{n,0}$ is the Kronecker delta.  We defined a sum
\be
\sum_{\{k\}} \equiv \prod_{l} \sum_{k_{l} \in \mathbb{Z}}\;,
\ee
over all configurations of the flux variables $k_{l}$. For magnetic matter we can analogously define $Z_m^{\, \lat}[U^m]$\footnote{Hoping that it will not cause confusion, we will keep the same notation for the magnetic current variable $k$, keeping in mind that it is a current on the dual lattice. The only place where this would need to be distinguished is when both magnetic as well as electric matter is written in a worldline formulation. While possible, none of our expressions are written in this form.}

\begin{align}\label{Zm_loop}
Z_m^{\, \lat} [U^m] & = \sum_{\{k\}} \, W_m[k] \; \prod_{\tilde x} \delta \big( (\vec{\nabla} \cdot \vec{k})_{\tilde x} \big) \; 
\prod_{\tilde l} e^{\, i \, A^m_{l} \, k_{\tilde l}} \; .
\end{align}

Each configuration comes with a weight factor
$W_e[k]$, which is specific for the theory under consideration. For the example systems which we discuss later 
the weights $W_e[k]$ are positive such that they are suitable for Monte Carlo simulations. The configurations of the 
flux variables are subject to the zero-divergence conditions at all sites, which in (\ref{Ze_loop}) 
we implement as the product of Kronecker deltas. Finally
the link variables are taken into account in the product of the $e^{i A_l k_{l}}$ at all links.

The next step is to rewrite the general gauge field Boltzmann factor $B_G^{\, \lat}[A]$ defined in (\ref{B_villain}). However, before we do that, let us first set the dimensionality to $d=4$ and consider a theory with one flavor of electric matter \eqref{S_Higgs} 
and one flavor of magnetic matter  \eqref{S_Higgs_m}. Recall that the magnetic gauge-field featured in the Boltzmann factor \eqref{B_villain} only as a linear factor in the action. Writing the magnetic matter \eqref{S_Higgs_m} in terms of the world-line representation \eqref{Ze_loop} we find that the total partition function with a single flavor of the magnetic matter has the form
\begin{align}
Z^{\, \lat}&= \int \!D[A^m]\int \!D[A]\prod_p \sum_{n_p\in\mathbb Z}\int D\phi\; e^{ - \sum_{ p}  \frac{\beta}{2} ( F_{ p}  + 2\pi n_{ p})^2-S_e[\phi,A]}\;,\cl
&\qquad\times\sum_{\{k\}} \, W_m[k] \; \prod_{\tilde \bx} \delta \big( (\star d \star k)_{\tilde \bx} \big)
\prod_{\tilde l} e^{\, i \, A_{\tilde l}^m \, k_{\tilde l}+ i A^m_{\tilde l}(\star d n)_{\tilde l} } \; .
\end{align}
Upon integrating over $A^m$, we find that $k_{\tilde l}=-(\star dn)_{\tilde l}$. Notice that this automatically satisfies the $(\delta k)_{ x}=0$ condition, because $d^2=0$. The final form of QED with one magnetic flavor is given by
\begin{align}\label{eq:wordline_monopole}
Z^{\, \lat}&=2\pi\int \!D[A]\prod_p \sum_{n_p\in\mathbb Z} e^{ - \sum_{ p}  \frac{\beta}{2} ( F_{ p}  +  2\pi n_{ p})^2  -S_e[\phi,A]} \, W_m[-(\star dn)] \;.\end{align}
In this form we have solved for all constraints, and as long as $W_m$ is a positive-definite weight we have managed to write a sign problem-free form of a theory with a single magnetic-matter field in $d=4$. We have thus found a lattice description of a theory of magnetic and electric charges which is completely sign problem free, at least as long as the magnetic matter is bosonic. We will soon see that such a theory can be dualized and the requirement is that either magnetic or electric matter is fermionic. We can also generalize the construction to $N_f$ flavors of magnetic matter. We have
\begin{align}
Z^{\, \lat}&= \int \!D[A^m]\int \!D[A]\prod_p \sum_{n_p\in\mathbb Z}\int D\phi\; e^{ - \sum_{ p}  \frac{\beta}{2} ( F_{ p}  +  2\pi n_{ p})^2-S_e[\phi,A]}\cl\;
&\qquad\times \sum_{\{k\}}W_{m}[k^{(1)},\dots,k^{(N_f)}]\prod_{f=1}^{N_f}\prod_{\tilde \bx} \delta \big( (\delta k^{(f)})_{\tilde \bx}  \big)\cl
&\qquad\times\prod_{\tilde l} e^{\, i \, A_{\tilde l}^m \, \sum_{f=1}^{N_f}k^{(f)}_{\tilde l}+ i A^m_{\tilde l}(\star d n)_{\tilde l} } \; ,
\end{align}
where the worldline factor $W_m$ now depends on the $U(1)$ currents of $N_f$ different monopole flavors, and where
\be
\sum_{\{k\}}\equiv\prod_{f=1}^{N_f}\prod_{\tilde l}\sum_{k_{\tilde l}^{(f)}\in \mathbb Z}\;.
\ee
Integrating over $A^m_{\tilde l}$ now introduces the constraint that $k^{(1)}_{\tilde l}+\dots k^{(N_f)}_{\tilde l}=-(\star dn)_{\tilde l}$, so we get, up to a constant factor,
\begin{align}
Z^{\, \lat}&=\int \!D[A]\prod_p \sum_{n_p\in\mathbb Z}\int D\phi\; e^{ - \sum_{ p}  \frac{\beta}{2} ( F_{ p}  +  2\pi n_{ p})^2-S_e[\phi,A]}\cl\;
&\qquad\times \sum_{\{k\}} \, W_{m_f}[k^{(1)},\dots,k^{(N_f)}]\prod_{f=1}^{N_f}\prod_{\tilde \bx} \delta \left( (\delta k^{(f)})_{\tilde \bx} \right)\cl
&\qquad\times \delta\left(k^{(1)}_{\tilde l}+\dots k^{(N_f)}_{\tilde l}+(\star dn)_{\tilde l}\right) \; .
\end{align}

Let us go back to the original gauge-theory in $d$ dimensions in Eq.~\eqref{B_villain}, and add electric matter in the form of world-lines \eqref{Ze_loop}. The Boltzmann factor \eqref{B_villain} is a product over all plaquettes $p$. At each plaquette we have a factor
\begin{align}\label{eq:poisson}
b\big(F_{p}, F_p^m\big) & =  \sum_{n_{p} \in \mathbb{Z}} 
e^{ - \frac{\beta}{2} ( F_{p}  +  2 \pi  n_{p})^2  -  i 
{(\star{F}^m)}_{p}  n_{p} }
\\
& =  \frac{1}{\sqrt{2 \pi  \beta}} \sum_{\tilde n^m_{p} \in \mathbb{Z}} e^{ - \frac{1}{2 \beta} \big( \tilde{n}^m_{p}  +  
\frac{({\star{F}^m)}_{p}}{2\pi} \big)^2 } \; e^{ i \tilde{n}^m_{p}  F_{p}+i\frac{1}{2\pi}(\star F^m)_{p}F_{p} } \; ,
\nonumber
\end{align}
which in the second line we have already expressed using the Fourier representation, i.e., the Poisson summation of Appendix B. Note that the final mixed term $(\star F^m) F$ will drop out upon summation over all plaquettes and partial integration\footnote{This is because $F=dA, F^m=dA^m$. Then by using the partial integration formula of Appendix~\ref{app:forms}  we have
\[
\sum_{p}(\star{F}^m)_pF_p=\sum_p(\star d F^m)_l A_l=0
\]
because $d^2=0$.}. 
However it would behoove us to notice that we can naturally define the Poisson-dual variables $\tilde n^m$ as integers living on the dual lattice. So let us define $\tilde n_p^{m}=(\star n^m)_p$, where $n^m$ is a $\mathbb Z$-valued $(d-2)$-form of the dual lattice.

We can therefore rewrite the product
of these factors over all plaquettes $p$ in terms of the dual plaquettes, which are $(d-2)$-cells $\tilde c^{(d-2)}=\star p$. The Boltzmann factor (\ref{B_villain}) can be rewritten as
\begin{align}
B_G^{\, \lat} [A,A^m] & =\prod_p b(F_p,\star F_p)=\cl
&=N_\beta \prod_{p}\sum_{n^m_{\star p}\in \mathbb Z}e^{-\frac{1}{2(2\pi)^2\beta}\left(F^m_{\star p}+2\pi n^m_{\star p}\right)^2+i n^m_{\star p}(\star F)_{\star p}}\;.
\label{BG_expand} 
\end{align}
In the second line we have introduced the normalization factor defined as $N_\beta = (1/\sqrt{2 \pi \, \beta})^{V d(d-1)/2}$, 
where in the exponent we have the total number of plaquettes given as the product of the number of sites 
$V = N_s^{d-1} N_t$ and the number $d(d-1)/2$ of plaquettes per site.

Now we want to integrate over the gauge field $A$ and eliminate the phase in the above formula. We write the phase as
\be
\sum_{p} n^m_{\star p}(\star dA)_{\star p}=\sum_{l}(\star dn^m)_lA_l \;.
\ee
Combining the above expression 
with the corresponding gauge field factor in the matter partition
function (\ref{Ze_loop}) and integrating over the measure (\ref{Ameasure}) we have
\be
\int D[A]\exp\left(i \sum_{l}\left[(\star d n^m)_l+k_l\right]A_l\right)=\prod_l\delta(k_l+(\star dn^m )_l)\;.
\ee
The integral over all gauge field configurations gives rise to a new constraint that sits on the links:
At every link $l$ the total flux generated by the variables $k_{l}$ for the matter field and by the 
 occupation numbers $\tilde n_{p}$ of the plaquettes $p$ whose link $l$ is in their boundary (i.e., for all $p\in\hat \partial l$) must vanish.

The final form of the dual representation of the partition function $Z^{\, \lat}$  then reads
\begin{align}
Z^{\, \lat}&=N_\beta  \int \!D[A^m]\prod_p\sum_{n^m_{\star p}\in \mathbb Z}e^{-\frac{1}{2(2\pi)^2\beta}\left(F_{\star p}^m+2\pi n^m_{\star p}\right)^2}\cl
&\qquad\times\sum_{\{k\}}W_e[k]\prod_{ x}\delta((\delta k)_{ x})\prod_{l}\delta(k_l+(\star dn^m)_l) \; .
\end{align}
Note that if we solve the second constraint $k_l=-(\star dn)_l$, the first one is automatically satisfied. So for one flavor of electric matter we can simply write
\begin{align}\label{eq:general_dual}
Z^{\, \lat}&=N_\beta\prod_p\int \!D[A^m]\sum_{n^m_{\star p}\in \mathbb Z}e^{-\frac{1}{2(2\pi)^2\beta}\left(F_{\star p}^m+2\pi n^m_{\star p}\right)^2}W_e[-(\star dn)] \; .
\end{align}
If we have $N_f$ flavors of electric matter then the expression is given by
\begin{align}\label{eq:general_dual_multiflavor}
Z^{\, \lat}&=N_\beta \int \!D[A^m]\int \!D[A]\prod_p\sum_{n^m_{\star p}\in \mathbb Z}e^{-\frac{1}{2(2\pi)^2\beta}\left(F_{\star p}^m+2\pi n^m_{\star p}\right)^2}\cl
&\qquad\times\sum_{\{k\}} \, W_{m_f}[k^{(1)}\dots k^{(N_f)}]\prod_{ \bx}  \prod_{f=1}^{N_f}\delta \big( (\delta k^{(f)})_{ \bx} \big) \big)\cl
&\qquad\times \delta(k^{(1)}_ l+\dots +k^{(N_f)}_{l}+(\star dn^m)_{ l}) \; ,
\end{align}
where 
\be\sum_{\{k\}}=\prod_{f=1}^{N_f}\sum_{\{k^{(f)\}}}\;.\ee

In $4d$ we can also couple the magnetic matter in a usual fashion, by just using the action \eqref{S_Higgs_m} for the magnetic matter (or its fermionic counterpart). We write explicitly only  the electric 1-flavor case, but the generalization to more flavors is straightforward
\begin{align}
Z^{\, \lat}&=N_\beta  \int \!D[A^m]\int D\tilde\phi\cl
&\qquad\times\prod_p\sum_{n^m_{\star p}\in \mathbb Z}e^{-\frac{1}{2(2\pi)^2\beta}\left(F_{\star p}^m+2\pi n^m_{\star p}\right)^2-S_m^{\, \lat}[\tilde \phi, A^m]}W_e[-(\star dn^m)] \; .
\end{align}

Notice that the similarity between the expression above and \eqref{eq:wordline_monopole} in $d=4$. This establishes a duality between $U(1)$ lattice gauge theory with electric and magnetic matter, and the same theory where electric and magnetic sectors are mapped into each other and  $\beta\rightarrow \frac{1}{(2\pi)^2\beta} $ (see a related discussion in \cite{Hands:1995ve}). This means that the lattice field theory with identical electric and magnetic matter  enjoys a self duality when $\beta=\frac{1}{2\pi}$. Furthermore we note that these theories can be formulated on the lattice without a sign problem as long as either magnetic or electric matter is bosonic. Specifically a theory with bosonic electric and magnetic matter has a self-dual point, and does not suffer from the complex-action problem.

It is perhaps useful to speculate on the phase diagram of a theory with equal number of electric and magnetic flavors of the same masses, such that at $\beta=1/{(2\pi)}$ the theory is self-dual. Since only bosonic matter is free of the sign problem in our formulation, we will restrict our attention to this case. We expect the phase diagram to be symmetric with respect to the singe change of the parameter $\log (2\pi\beta)$ which changes its sign under the duality transformation. Further we can distinguish four phases: the photon phase where neither electric nor magnetic matter condenses, the (dual-)superconductor phase where electric (magnetic) matter condenses, and the mixed phase where both electric and magnetic matter condenses. The last alternative is a strange one, as it is not expected to appear in the continuum. This is because a superconducting phase has monopole confinement, and the dual-superconductor phase has electric confinement. The confining strings are ANO vortices for which, depending on the phase, one of the scalar field's (electric or magnetic) phase has nontrivial winding around a worldsheet of the vortex. In the continuum this requires the scalar field to go to zero in the center, endowing the vortex with tension, and confining electric (magnetic) charges when magnetic (electric) condensation occurs. This argument may break down on a finite lattice for some parameters which compress the string to the lattice-spacing size, allowing a condensation of both electric and magnetic charges.

The mixed phase would be expected when the  mass parameters $m^2_e$ and $m^2_{m}$ are negative and large compared to the inverse lattice spacing squared. For less extreme, but still negative values of $m^2_{e,m}$ we expect a superconductor/dual-superconductor phase transition as a function of $\beta$, which for identical parameters of electric and magnetic matter will be exactly at the point when $\beta=1/({2\pi})$, likely in the form of a first order phase transition. This phase transition line should end with a critical, second order phase transition to the photon phase for even larger values of $m^2_e=m^2_m=m^2$. 

We note that the magnetic and electric superconductor phases may be hard to distinguish in case of the single flavor unit-charge electric and magnetic matter, because there is no symmetry in the problem other than the electric-magnetic duality which is only present when $\beta=1/(2\pi)$. It may therefore be more precise to consider a theory with charge-2 magnetic and electric matter, so that there is a $\mathbb Z_2\times \mathbb Z_2$ electric-magnetic center symmetry. The superconductor and the dual-superconductor can then be defined as phases for which non-contractible Wilson loops for $A$ and $A^m$ get a vev respectively, breaking either one $\mathbb Z_2$ or the other. In fact, using arguments similar to the ones in \cite{Komargodski:2017smk}, one can show that there is a mixed anomaly between the two center symmetries, and the gapped phase where both are restored is not permissible. Alternatively we can also consider a theory with $n_f^e$ and $n_f^m$ flavors of electric and magnetic matter respectively in such a way that they form $SU(n_f^m)$ and $SU(n_f^e)$ multiplets. The electric/magnetic condensation can then be associated with spontaneous breaking of $PSU(n_f^m)$ and $PSU(n_f^e)$ global symmetries\footnote{The global symmetries are $PSU(N)$ instead of $SU(N)$ because the transformation by a center of $SU(N)$ is a gauge symmetry.}.

We conclude by commenting on the theories with fermions. 
For fermionic matter a positive-definite form of the wordlines is not so simple to achieve, but at least in $2d$ 
some fermionic systems were shown to have positive weights $W_e[k]$ (see e.g., 
\cite{Gattringer:2015nea,Gattringer:2015cxh,Goschl:2017kml}). This implies that the wordline/worldsheet representation
can be simulated with Monte Carlo techniques and for a discussion of suitable updating techniques we refer to 
\cite{Mercado:2013yta}. However as we have seen, the theory only requires a world-line representation of magnetic or electric matter, but not both. This means that as long as either magnetic or electric sector is purely bosonic, the theory can be formulated in a representation that does not suffer from a sign-problem.

\section{\texorpdfstring{$2d$  $U(1)$}{2d U(1)} gauge theories with a \texorpdfstring{$\theta$}{theta}-term}
\label{sectionGHtheta}

As a first example of our approach with mixed compact/non-compact lattice fields we discuss two 
$2d$ abelian gauge models. One is the Higgs model with a $\theta$-term, and the other one is the 
$CP(N-1)$ model with the $\theta$ term. The former can easily be generalized to arbitrary number of 
flavors, but to keep formulas simple we here discuss only one flavor. For a proper understanding of the 
model from lattice simulations it is necessary to discretize the $\theta$-term such that it is integer-valued 
and has a form that is suitable for solving the complex action problem by a transformation to worldlines 
and worldsheets. Both these conditions are met by our discretization approach and as a consequence the 
lattice model at non-trivial topological angle $\theta = \pi$ has an exact charge conjugation symmetry.  

In \cite{Gattringer:2018dlw} we already numerically studied the 1-flavor gauge Higgs model. This system 
undergoes a spontaneous breaking of charge-conjugation symmetry as a function of the mass parameter and  
the transition is in the $2d$ Ising universality 
class (see also \cite{Komargodski:2017dmc}), thus demonstrating the power of our approach.
When switching to charge-2 matter fields or more than one flavors of matter one expects that the critical behavior 
changes, and currently we are exploring the corresponding physics (see \cite{Goschl:2018uma} for our 
preliminary results).

\subsection{The models and their duals}

In the continuum the actions of the $2d$ models we  consider are
\begin{equation}
S_{\text{Higgs}}[\phi,A] = \int d^2x\;\Big(\frac{1}{4e^2}F_{\mu\nu}^2+ |D_\mu \phi(x)|^2 + m^2 |\phi(x)|^2 + \lambda |\phi(x)|^4 \Big) \;,
\end{equation}
and
\begin{equation}\label{eq:CPN_action}
S_{CP(N-1)}[\phi,A] = J\sum_{j=1}^N|D_\mu z_j(x)|^2\;,
\end{equation}
where $z_j(x), j = 1,2 \, ... \, N$ are $N$ species of complex scalars with the constraint 
$\sum_{j=1}^Nz_j(x)^*z_j(x)=1\; \forall x$. The covariant derivative in the $CP(N-1)$ case contains a 
non-dynamical auxiliary gauge field which can be integrated out because there is no gauge-kinetic term. 
Both theories can also be supplemented with a topological $\theta$-term which in the continuum is given by 
($\epsilon_{\mu \nu}$ denotes the completely anti-symmetric tensor)
\begin{equation}
S_\theta=\frac{\theta}{4 \pi} \int \!\!d^{\,2}x \; \epsilon_{\mu \nu} \, F_{\mu \nu}(x) \; = \; \frac{\theta}{2 \pi} \int \!\!d^{\,2}x \; F_{12}(x) \; .
\label{theta_cont}
\end{equation}
The lattice discretization would involve a term $\frac{i\theta}{2\pi}\sum_p B_p$, where $B_p=2\pi n_p$ is  $B_p$ is the $\mathbb Z$ gauge field that we used extensively. In fact the construction is identical to that in general dimensions upon replacement of $(\star F^m)_{p}\rightarrow \theta$. Indeed in the first line of Eq.~\eqref{eq:poisson} this replacement amounts to adding the $\theta$-term in $2d$, using the prescription above. 

However, this will only work if the gauge-kinetic term is present, which is not the case in the continuum version of the $CP(N-1)$ model. The naive discretization of the continuum formulation without the gauge-kinetic term will yield a meaningless result, rendering a   vanishing partition function for any value of $\theta$ except $0\bmod\; 2\pi$. This is because the 2-form variable $B\in 2\pi \mathbb Z$, or equivalently the Villain variable $n\in \mathbb Z$, decouples from the theory. Since the $\theta$-term couples only to $B$, the sum over $B$ will force the partition function to vanish, a clearly unphysical result. 

The problem comes from the ill-defined topology of the lattice model. In fact without the gauge-kinetic Villain term there is no penalty for instantons at the lattice scale (so called ''defects''), and the theory will not have a proper continuum limit. On the other hand the introduction of the gauge-kinetic term is harmless, and indeed necessary, as long as $\beta$ is held constant as the continuum limit is performed. This can be seen in several ways. One is to observe that on dimensional grounds the gauge-kinetic term must be proportional to $a^2 \beta\int d^2x F^2$. Since $\beta$ is held constant as $a\rightarrow 0$, the term vanishes in the continuum limit, which is attained by sending $J\rightarrow \infty$ in this model. 

Another way to see that indeed the term will do no harm in the continuum limit is to consider the physics of this term. Physically the term will induce a Coulomb interaction between oppositely charged particles. Since we are in one spatial dimension, the Coulomb interaction is confining, and the typical worldline separation of oppositely charged wordlines in the vacuum will be given by a distance $\sim \beta$ in lattice units. In the limit of a vanishing gauge-kinetic term, all $U(1)$ charged worldlines collapse to only $U(1)$ gauge-neutral wordlines. However, if a fixed $\beta$ is imposed on the lattice, this will fix the typical distance between $U(1)$ charged wordlines to $\sim \beta$ in lattice units. Since any finite length in lattice units collapses to zero in the continuum limit, both the theory with and without kinetic term have the feature of suppressing wordlines with a $U(1)$ charge, with the added bonus that the kinetic term smoothens out gauge-field configurations at the lattice scale. So in what follows we will always endow the $CP(N-1)$ model with a gauge-kinetic term, keeping in mind that its coupling will be held fixed in the continuum limit.

Upon replacement $(\star F^m)_p\rightarrow \theta$ we find from the general dual expression \eqref{eq:general_dual} that the model with one flavor is given by
\begin{align}\label{eq:2d_dual}
Z^{\, \lat}=&\sum_{n^m_{\tilde x}\in \mathbb Z}e^{-\frac{1}{2(2\pi)^2\beta}\left(2\pi n^m_{\tilde x}+\theta\right)^2}W_e[k]\prod_l\delta(k_l+(\star d n^m)_l)\prod_x\delta((\delta k)_x)\cl
&=N_\beta\prod_{\tilde x}\sum_{n^m_{\tilde x}\in \mathbb Z}e^{-\frac{1}{2(2\pi)^2\beta}\left(2\pi n^m_{\tilde x}+\theta\right)^2}W_e[-(\star dn^m)]\;.
\end{align}
where in the last step we solved for the constraint $k_l=-(\star dn^m)_l$. 

In the case of the $CP(N-1)$ model we have $N$ flavors that generate $N$ different $U(1)$ currents, so we must do the replacement $(\star F^m)_p\rightarrow \theta$ in \eqref{eq:general_dual_multiflavor} and find
\begin{align}
Z^{\, \lat}&=N_\beta\prod_{\tilde x}\sum_{n^m_{\tilde x}\in \mathbb Z}e^{-\frac{1}{2(2\pi)^2\beta}\left(2\pi n^m_{\tilde x}+\theta\right)^2}\cl
&\qquad\times\sum_{\{k\}} \, W_{m_f}[k^{(1)}\dots k^{(N)}]\prod_{ \bx}  \prod_{f=1}^{N}\delta \big( (\delta k^{(f)})_{ \bx} \big)\cl
&\qquad\times \prod_l\delta(k^{(1)}_ l+\dots +k^{(N)}_{l}+(\star dn^m)_{ l})
\end{align}

Note that at $\theta=\pi$ the models enjoy a charge-conjugation symmetry which takes $n^m_{\tilde x}\rightarrow -n^m_{\tilde x}-1$, as long as the matter field weights obey $W_e[k]=W_{e}[-k]$. All that is left is to find the expressions 
for $W_e[k]$ in the case of specific models. We will discuss these matter field weights 
for the gauge Higgs model and the $CP(N-1)$ model.

\subsection{Worldlines for the mater field weights}

Now we state the worldline expressions for the two models we study. As derived in Appendix \ref{app:Higgs}  
the worldline expression for the gauge-Higgs model is
\begin{eqnarray}
W_e[k] &\!\!\!\!=\!\!\! \!& \sum_{\{a\}} \left[\prod_{x,\mu} \frac{1}{\left(|k_{x,\mu}|\!+\!a_{x,\mu}\right)! \ a_{x,\mu}!} \right]
\left[\prod_x \! I\left(f_x\right)\right] \qquad \mbox{with}
\label{W_e_$2d$GH} \\ 
I\!\left(f_x\right)  & \!\!\! \! = \!\!\! \!&  \! \!\int_{0}^{\infty} \!\!\!\!\!\! dr\, r^{f_x+1} e^{-M r^2 -\lambda r^4}\!\! , \;\, 
f_x   = \!\! \! \ \sum_{\mu}
\big[ |k_{x,\mu}| \!+\! |k_{x-\hat{\mu},\mu}| + 2\left( a_{x,\mu} \!+\! a_{x-\hat{\mu},\mu} \right) \big] .
\nonumber
\end{eqnarray}
For the $CP(N-1)$ model we have (see Appendix \ref{app:CPN})
\begin{align}
&W_e[ k^{(1)}\!\!,\dots,\!k^{(N)}]   =  \sum_{\{a\}}  \prod_{x,\mu,j} 
\frac{ J^{ | k_{x,\mu}^{(j)} |  +  2 a_{x,\mu}^{(j)} } } 
{ ( | k_{x,\mu}^{(j)} |  +  a_{x,\mu}^{(j)} )! \; a_{x,\mu}^{(j)} ! } 
\prod_x   \frac{\prod_{j=1}^N \left( { f_x^{(j)}  / 2} \right) ! }{
\left( N-1+ \frac{1}{2}  \sum_{j=1}^N  f_x^{(j)}  \right) ! } \cl
&\qquad \mbox{with}
\cl 
& f_{x}^{(j)}   =   
 \sum_\mu \Big[ |k_{x,\mu}^{(j)}| + |k_{x-\hat{\mu},\mu}^{(j)}| + 2[a_{x,\mu}^{(j)} + a_{x-\hat{\mu},\mu}^{(j)}] \Big] 
 \; , \;j = 1,2 \, ... \, N \; ,
\label{cpn_WM}
\end{align}
where $a_{x,\mu}\in\mathbb N_0$ in (\ref{W_e_$2d$GH}) and $a_{x,\mu}^{(j)} \in\mathbb N_0$
in (\ref{cpn_WM}) are auxiliary link variables on the original lattice.

\subsection{Some comments on gauge choices and the \texorpdfstring{$\theta$}{theta}-term in \texorpdfstring{$2d$}{2d}}

In this section we will discuss some specific gauge choices in $2d$ and make contact with more traditional 
definitions of the
$\theta$-term. However, before we begin let us note that the most obvious gauge in any dimension is the Villain 
gauge, which sets the link gauge-fields $A_l$ to be in the interval $A_l\in [-\pi,\pi)$ for every link $l$ of the lattice $
\Lambda$. This is simple because we can always perform a discrete 1-form gauge transformation at each link 
individually. 

Now we want to show that the $\theta$-term we suggested in (\ref{eq:action_theta}) really is the $\theta$-term by connecting 
it with conventional formulations.  To do this let us first open up the lattice 2-torus, and put it on a grid of size 
$N_t\times N_s$. Normally, for implementing toroidal boundary conditions,  
we identify each vertex and link on the far-left with the corresponding vertex and link on the right, and the same with 
the vertices/links on the upper/lower edge of the grid. However, let us for the moment drop this condition. In this case 
we can gauge away the field $B_p$ completely, by just setting it equal to a total derivative $B_p=(d\lambda)_p$, 
where $\lambda_l\in 2\pi \mathbb Z$ is a link-valued field. This is indeed possible because we can
choose $\lambda_l = 0$ for all links $l$ that are horizontal 
and then set the vertical links to be equal to the sum of all $B_p$ on the plaquettes to the left of the link. 
We can then perform the gauge transformation \eqref{eq:1-form_gauge_transform} and gauge away $B_p$ 
completely. However, now the gauge parameter on the opposing vertical links which are to be identified are 
not the same, which means that the gauge fields $A_l$ are no longer the same on the opposing links, but differ by a 
value in $2\pi\mathbb Z$. This indeed makes sense because now the  gauge invariant combination $\frac{\theta}{2\pi}
\sum_p(F_p+B_p)$ needs to be used as the $\theta$-term, and since we have set all $B_p$ to zero, the contribution to 
the $\theta$ term must be coming from the edges of the grid. The gauge field $A_l$ can now be thought of as a 
$U(1)$ gauge field, since it is not periodic on the torus, but only periodic up to a $U(1)$ gauge transformation.

Another way to make the connection with the $\theta$-term is to define the $U(1)$ gauge theory in the 
conventional way, i.e., with the Wilson action for compact $U(1)$ lattice gauge fields and a naive discretization of the 
topological charge. Using $F_p = (dA)_p$  the action with the $\theta$ term reads
\be
S \; = \; \sum_p\left(-\beta\cos(F_p)+i\frac{\theta}{2\pi} \sin(F_p)\right)\;.
\label{SG_theta_conventional}
\ee
In the conventional representation the link field $A_l$ defines a phase on links, such that it assumes values  
$A_l\in[-\pi,\pi)$. However, we can redundantly assume that $A_l\in(-\infty,\infty)$ and in this way have an 
explicit gauge invariance $A_l\rightarrow A_l+2\pi n_l, n_l\in\mathbb Z$. Now if we take $\beta$ to be large, we can 
approximate the trigonometric functions in (\ref{SG_theta_conventional}) by their expansion around the minimum of 
$-\cos(F_p)$. However there are infinitely many minima of the cosine, and we must sum over them. 
In this way we obtain (constants were dropped)
\be
S \; = \; \sum_{p}\left(\frac{\beta}{2}(F_p+2\pi n_p)^2+i\frac{\theta}{2\pi}(F_p+2\pi n_p)\right)+O((F_p+2\pi n_p)^3)\;,
\ee
where $n_p\in \mathbb Z$ labels the minimum of the $\cos F_p$ around which we expand. Since  $F_p$ is a 
total lattice derivative, it will drop in the linear term under the sum. By identifying $B_p=2\pi n_p$, we get back the 
action \eqref{eq:gauge_action_B}, with the $\theta$-term \eqref{eq:action_theta}.

It is instructive to discuss another gauge choice in $2d$ where we can remove the field $B_p$ everywhere 
except on a single plaquette. For this we use (\ref{Bp_gaugetrafo}) which states that under the transformation
$A_l \rightarrow A_l+ \lambda_l$ with $\lambda_l = 2\pi n_l$ and $n_l \in \mathbb{Z}$ the field $B_p$ transforms as  
$B_p \rightarrow  B_p - (d \lambda)_p$. For defining the gauge transformation we switch to a notation where we 
denote sites as $x = (x_1,x_2)$, the link fields $\lambda_l$ as $\lambda(x_1,x_2)_\mu$ and the plaquette-based
fields $B_p$ as $B(x_1,x_2)$, where we again make use of the fact that in 2$d$ plaquettes can be labelled by 
their lower left corner $x = (x_1,x_2)$.

The gauge transformation proceeds in two steps. For a first gauge transformation we define
\begin{equation}
\lambda(x_1,x_2)_1 \, = \, 0 \; , \quad 
\lambda(x_1,x_2)_2 \, = \, \sum_{t=1}^{x_2-1} B(x_1,t) \; .
\end{equation}
The transformed field $B'_{p} = B_{p} - (d\lambda)_p$ is zero everywhere except on the upper strip of the open 
torus. For the second step we define
\begin{equation}
\lambda(x_1,x_2)_1 \, = \, \sum_{s=1}^{x_1-1} B'(s,x_2) \; , \quad
\lambda(x_1,x_2)_2 \, = \, 0 \; ,
\end{equation}
and obtain for the transformed field $B''_p = B'_p - (d\lambda')_p$ a field that vanishes everywhere except 
for the plaquette at $(x_1,x_2) = (N_s,N_t)$. So by a gauge transformation we were able to set the field $B_p$
to zero everywhere except on a single plaquette. The value of the field $B_p$ on this plaquette 
determines the $\theta$-dependence of the system. 

Note, however, that this gauge choice does not specify the gauge completely. To see this, imagine an arbitrary 
closed contour on the links of the dual lattice. Let $\{l\}$ be the set of all links $l$ of the original lattice which 
intersect the contour. Now we implement a $\mathbb{Z}$-valued 1-form gauge transformation on all links $l\in\{l\}$, 
i.e., $A_l\rightarrow A_l+2\pi m_l$, where $m_l=m_{l'}$ if $l,l'\in\{l\}$, and $m_l=0$ if $l\notin\{l\}$. This clearly does not 
change $B_p$. 

We remark at this point that when using this particular gauge for the mapping to dual variables, which we discuss
in the next section, gives rise to the same worldline and worldsheet representation we derive below.

Finally let us make contact with the geometrical topological charge \cite{Berg:1981er,Luscher:1981zq}. To do this let us also revert back to the usual formulation of the abelian gauge theories in terms of lattice $U(1)$-valued link fields $U_l=e^{i A_l}$, and redundantly integrate over $A_l\in \mathbb R$. Clearly this just adds an infinite multiplicative factor which is a gauge symmetry artifact because the transformation $A_l\rightarrow A_l+2\pi$ is a gauge symmetry of the theory. 
Note, however, that this formulation does not have an associated 1-form gauge field $B_p$. We can define the field strength in two ways. One way is to define it as before and set $F_p=(dA)_p$. The other way we can define it through $V_p$ as $V_p=e^{i f_p}$, in which case $f_p$ is only defined $\bmod 2\pi$. Note that $F_p$ is clearly not invariant under the 1-form gauge symmetry, as it can shift by $2\pi$. The definition of $f_p$ on the other hand is ambiguous and a prescription of what it means is required before it can be used to define the $\theta$-angle. The geometrical prescription is to define the $\theta$ term using $f_p$ in such a way that it is picked to satisfy $-\pi\le f_p<\pi$, and we write the 
$\theta$-term as $S_{\theta}=\frac{\theta}{2\pi}\sum_p f_p$. 

We want to connect this definition with our definition given in terms of the $B$-field. To define the $\theta$-term in that  
way we will need to first define an adequate $B_p$ field. 
We proceed as follows: we first insert unity for every plaquette $p$ into the partition function in the form
\be
\sum_{n_p=-\infty}^\infty \Theta(F_p+2\pi n_p)\;,
\ee
where $\Theta(x)$ is a Heaviside-like function defined as
\be
\Theta(x)=\begin{cases}
1 & -\pi\le x< \pi\\
0 & \text{otherwise}
\end{cases}\;.
\ee
Note that we have not changed the partition function at all by this introduction. What we have done is to partition $F_p$ into 
individual segments in the interval $-\pi-2\pi n_p\le F_p<\pi-2\pi n_p$. However, 
the combination $2\pi n_p$ now behaves as the gauge field $B_p$. 

Notice that the combination $F_p+2\pi n_p$ can be identified with $f_p$. This would exactly correspond to taking the 
combination $F_p+2\pi n_p$ for the $\theta$-term, as per L\"uscher's prescription \cite{Berg:1981er,Luscher:1981zq}. 
However $F_p$ is an exact 2-form, and will drop out under the sum, rendering the L\"uscher $\theta$-term equivalent to 
our definition: $Q=\frac{1}{2\pi}\sum_p(F_p+2\pi n_p)=\sum_p n_p$.

\section{\texorpdfstring{$3d$ $U(1)$}{3d U(1)} gauge theory with constrained monopole charge}
\label{sectionGHmonopoles}

The next system we consider is the $3d$ $U(1)$ gauge theory with a general monopole charge. Recall that in the absence of monopoles -- or magnetic flux changing events -- the theory has an ordinary $U(1)$ symmetry associated with  magnetic flux conservation, sometimes referred to as the $U(1)$-topological symmetry which we will refer to as $U(1)^T$. If the monopole charge is restricted to be a multiple of some integer $N$, then $U(1)^T$ is reduced to $\mathbb Z_{N}^T$. In Sec. \ref{discretization} we already touched on how the monopole charges can be restricted. 

Such a setup is interesting in several contexts. Firstly, various non-abelian gauge theories in $3d$ and $4d$ have regimes where the effective theory is an abelian gauge theory with monopoles. In some of them, on symmetry grounds, odd monopole charges are forbidden.

Secondly,  $(2+1)d$ spin-1/2 anti-ferromagnets may support a Valence Bond Solid (VBS) phase -- a phase which can be visualized as a pairing of spins into singlet dimers -- which is also effectively described by an abelian gauge Higgs theory with two scalars (see, e.g., \cite{Murthy:1989ps,senthil2004a,senthil2004b,Sachdev:2015slk} and references therein). The lattice symmetries of these spin systems map into the topological symmetry $\mathbb Z_{N}^T$ discussed above, which in turn restricts the monopole charges of the effective theory. The effective theory has been used to argue that the transition from the VBS phase (i.e., monopole phase) to the N\'eel phase (i.e., Higgs phase) is a second order quantum phase transition. 

In this section we will briefly discuss how to formulate the theory with restricted monopole charges on the lattice and subsequently show how the complex action problem is overcome by switching to the worldline/worldsheet representation.

\subsection{Defining the constraints for the monopole charge}

We begin with the definition of the monopole charge density $q(x)$ at a space-time point $x$ in the continuum, 
which is simply given by the divergence of the field components in the field strength tensor,
\begin{equation}
q(x) \; = \;  \, \frac{1}{4\pi}  
\sum_\sigma \, \partial_\sigma \, \sum_{\mu, \nu} \, \epsilon_{\sigma \mu \nu}  \, F_{\mu \nu}(x) \; .
\label{qdefcont}
\end{equation}
We already briefly discussed monopoles on the lattice  -- see the discussion around Eq.~\eqref{eq:Monopole_Action}. The prescription there can also be thought of in terms of a monopole charge density $q_x$ that corresponds to (\ref{qdefcont})
by discretizing the derivative $\partial_\sigma$ 
and implementing the prescription (\ref{eq:1-form_gauge_transform}). 
The result is the monopole charge density $q_x$ assigned to 
the cubes of the lattice (i.e.,~the dual lattice site), which we label by the dual-site $\tilde x$:
\begin{equation}
q_{\tilde x}[n] \; = \frac{1}{2\pi} \sum_{\sigma} \sum_{\mu < \nu}   \epsilon_{  \sigma \mu \nu}  
\Big[ ( n_{x+\hat{\sigma},\mu \nu} \; - \; n_{x,\mu \nu} )\, \Big] \; .
\label{qdef_lat}
\end{equation}
Note that here we sum over $\mu < \nu$ which gives rise to an extra factor of 2 and thus to 
a normalization with $1/2\pi$. The definition above is consistent with our discussion around Eq.~\eqref{eq:Bflux_constraint}.

The constraint for the monopole charge assigned to the cube at $x$ is imposed with a Lagrange multiplier as
\begin{equation}
\frac{1}{N} \sum_{a_{\tilde x} = 0}^{N-1} \, e^{\, i \frac{2 \pi}{N} \, a_{\tilde x} \, q_{\tilde x}[n]}  \; .
\label{cubeconstraint}
\end{equation}
We see that this is the same as setting $A^m_{\tilde x}=\frac{2\pi a_{\tilde x}}{N}$. We can therefore easily write the dual version by replacing $A^m_{\tilde x}\rightarrow \frac{2\pi}{N} a_{\tilde x}$ in the expression, and converting the integral over $A^m_{\tilde x}$ into a sum over $a_{\tilde x}$. We therefore write the expression for the dual representation of the partition function by the appropriate replacement in Eq.~\eqref{eq:general_dual}. For one flavor we have
\begin{align}
Z^{\, \lat}&=\prod_{\tilde x}\sum_{a_{\tilde x}=0}^{N-1}\prod_{\tilde l}\sum_{n^m_{\tilde l}\in \mathbb Z}e^{-\frac{1}{2\beta}\left(\frac{(da)_{\tilde l}}{N}+n^m_{\tilde l}\right)^2}W_e[-(\star dn)]\;,
\end{align}
We note two things: Firstly, the above construction can be seen as a limiting case of a $N$-clock deformation of the theory, i.e.,~by adding the monopole action \eqref{eq:Monopole_Action} and setting $m_q=0$ for all $q\ne0$ and sending $m_N\rightarrow +\infty$. This potential precisely pins down the $A^m$ field to be $(A^m)_{\tilde x}=2\pi a_{\tilde x}/N$ with $a_{\tilde x}\in \mathbb Z$.    Secondly, the whole construction can easily be repeated for the general monopole action defined in \eqref{eq:Monopole_Action}, by simply writing the cosines as sums of exponentials, and applying our standard construction. Note that if some of the coefficients $m_n$ in \eqref{eq:Monopole_Action} are negative, we might face sign problems. 

All that is left is to write down the expression for the worldlines $W[k]$. In fact this expression is identical to the one in $2d$. We write it out again for convenience
\begin{eqnarray}
W_e[k] &\!\!\!\!=\!\!\! \!& \sum_{\{a\}} \left[\prod_{x,\mu} \frac{1}{\left(|k_{x,\mu}|\!+\!a_{x,\mu}\right)! \ a_{x,\mu}!} \right]
\left[\prod_x \! I\left(f_x\right)\right] \qquad \mbox{with}
\label{W_e_$3d$GH} \\ 
I\!\left(f_x\right)  & \!\!\! \! = \!\!\! \!&  \! \!\int_{0}^{\infty} \!\!\!\!\!\! dr\, r^{f_x+1} e^{-M r^2 -\lambda r^4}\!\! , \;\, 
f_x   = \!\! \! \ \sum_{\mu}
\big[ |k_{x,\mu}| \!+\! |k_{x-\hat{\mu},\mu}| + 2\left( a_{x,\mu} \!+\! a_{x-\hat{\mu},\mu} \right) \big] .
\nonumber
\end{eqnarray}

\section{The \texorpdfstring{$\theta$}{theta}-terms in four dimensions and the Witten effect}\label{sec:4d}

{In this section we will discuss some $\theta$-terms in four-dimensional gauge theories which arise naturally from our discussion in the rest of the paper. We discuss two cases: the abelian $U(1)$ gauge theory in $4d$ and the $SU(N)/\mathbb Z_N$ gauge theory. The basic idea is along the spirit of \cite{Kapustin:2014gua}, where non-abelian $SU(N)/\mathbb Z_N=PSU(N)$ gauge theory was discussed. In \cite{Kapustin:2014gua} a deformation of such theories by a discrete $\theta$-term was discussed in the continuum and on manifold triangulations (i.e., simplicial complexes). To our knowledge no construction of these $\theta$-terms was presented for cubic lattices. We also are unaware of any discussion along these lines for abelian gauge theories either on triangulations or cubic lattices. 

We note that this section discusses cases which are manifestly suffering from the complex-action problem, and we will not try to dualize them here. This seems to be possible however, but is beyond the scope of this work and will be discussed elsewhere\footnote{We are thankful to Bruno Le Floch for discussions on this topic}.

\subsection{The \texorpdfstring{$U(1)$}{U(1)} gauge theory}

In $d=4$, and in the absence of magnetic matter, we can write a natural $\theta$-term, as we will discuss momentarily. 
However, the construction needs slightly more structure. Namely since the cubic lattice $\Lambda$ has as its dual lattice also a cubic lattice $\tilde\Lambda$, there is a natural map $F:\tilde\Lambda\rightarrow \Lambda$ which performs a pure translation  $\tilde x\rightarrow x$. Furthermore this map also induces a map between fields of the dual lattice and the original lattice. We will therefore identify the two lattices as equivalent, having the same field content. This allows us to define the operator $\star_F= F\circ \star$, which maps $r$ forms on $\Lambda$ to $d-r$ forms on $\Lambda$. Note, however, that the square $(\star_F)^2 c^{(r)}=F(\star F(\star c^{(r)}))=(-1)^{r(d-r)}F^2(c^{{(r)}})$ does not leave the cell invariant, but instead translates it by one lattice unit in all (negative) directions. This operator should be thought of as the Hodge dual operator on the lattice, but note that, unlike the Hodge dual, it also translates the lattice. This will be important below.

Let us therefore define the fields $\tilde B_{\star p}\equiv B_{F(\star p)}$ and $\tilde A_{\star l}\equiv A_{F(\star l)}$. We  can now write a $\theta$-term for U(1) gauge theory in 4$d$ as
\be
S_{\theta}^{4d}=\frac{i\theta}{8\pi^2}\sum_{p}B_{p}\tilde B_{\star p}\;.
\ee
When we impose the condition $dB=0$ then the above term is invariant under 
$\theta\rightarrow \theta+2\pi$ if $dB=0$, as we will show in a moment.


First, let us make sure that the $\theta$-term is gauge invariant under the $1$-form gauge symmetry $B_p\rightarrow B_{p}+2\pi (dn)_p$, where $n_l$ is a gauge variation, as long as $(dB)_{c}=0,\forall c$. Indeed
\begin{multline}
\Delta S_{\theta}^{4d}=i\frac{\theta}{4\pi}\sum_p\left((dn)_p \tilde B_{\star p}+B_{p}(d\tilde n)_{\star p}+2\pi (dn)_p(d\tilde n)_{\star p}\right)=\\
=-i\frac{\theta}{4\pi}\left(\sum_{l}n_l (d\tilde B)_{\star l}+\sum_{\tilde l} (dB)_{\star \tilde l} \tilde n_{\tilde l}\right)=\\
=-i\frac{\theta}{4\pi}\left(\sum_{l}(dB)_{F^2(\star \tilde l)}\left(\tilde n_{\tilde l}+\tilde n_{F^{-2}( \tilde l)}\right)\right) \; ,
\end{multline}
where in the last step we used the partial integration theorem of Appendix \ref{app:forms}. Indeed the above variation vanishes identically if $dB=0$. 

Now we show that the term is also invariant under a  shift of $\theta\rightarrow \theta+2\pi$. This shift produces a term
\be
e^{i\pi \sum_{p}b_p\tilde b_{\star p}} \; ,
\ee
where we wrote $B_p=2\pi b_p$ with $b_p\in \mathbb Z$.  The individual contributions from the plaquettes can indeed be odd. We must therefore show that the number of odd contributions is even. Let's assign for every odd $b_p$ an integer $k_{\star p}=1$ on the dual lattice, and for every even $b_p$ and integer $k_{\star p}=0$. Then because we have $(db)_c=0$, we must have that $(\delta k)_{\tilde l}=0$, in other words that $k=1$ two-dimensional surfaces on the dual lattice are closed surfaces. The same is true for $\tilde b_{\tilde p}$ on the dual lattice, to which $\tilde k_{\star \tilde p}=0$ or $\tilde k_{\star \tilde p}=1$ on the original lattice depending on whether $\tilde b_{p}$ is even or odd respectively. Then each plaquette on the original lattice contributes a term
\be
e^{i\pi k_{\star p} \tilde k_p}\;.
\ee
The above term is $-1$ whenever the  $k=1$ surface on the dual lattice intersects the $\tilde k=1$ surface on the original lattice, but is $1$ otherwise. However, because $k$ and $\tilde k$ are related by a simple translation of the lattice to the dual lattice $F:\Lambda\rightarrow \tilde\Lambda$, there is a doubling of the intersection points, and the product over plaquettes yields unity\footnote{By a version of this argument one can see that the $\theta$-term is really a cup product $\frac{B}{2\pi}\cup \frac{B}{2\pi}\in H^4(\mathcal M,\mathbb Z)$. The argument uses the $\star$-operator to construct closed surfaces on the dual lattice and the original lattice as before, and shows that the contribution is the intersection number of these surfaces, which is Poincare dual to the cup-square of $B\in H^2(\mathcal M, \mathbb Z)$.}.

So far all the discussion was without monopoles. To couple monopoles we replace the constraint $(dB)=0$ with a Lagrange multiplier by inserting the term \eqref{eq:deformation} which can also be written as
\be
i\Q=\frac{i}{2\pi}\sum_{\tilde l}A^m_{\tilde l} (dB)_{\star \tilde l}\;.
\ee

Now we are forced to define the $1$-form gauge symmetry as a simultaneous transformation
\begin{align}
&A_l\rightarrow A_l+2\pi n_l\cl
&B_p\rightarrow B_p+2\pi (dn)_p\\
&A^m_{\tilde l}\rightarrow A_{\tilde l}^m+ \frac{\theta}{2}\left( \tilde n_{\tilde l}+\tilde n_{F^{-2}(\tilde l)}\right)\;.\nonumber
\end{align}
Note that in order for the formulas to make sense for general $\theta$, we must take $A_{\tilde l}^m\in \mathbb R$. If we restrict the interval $A^m_{\tilde l}\in[0,2\pi)$ then the last replacement  should be thought of as a $\bmod \; 2\pi$ statement.

The expression above makes it clear that in order to couple the magnetic monopoles with the above theory, one must endow them with electric charge. Note, however, that the coupling of monopoles to electric fields is unusual, as the lattice hopping term must contain a combination $A^m_{\tilde l}-\frac{\theta}{4\pi}\left(A_{\tilde l}+A_{F^{-2}(\tilde l)}\right)$. This is a consequence of the fact that $\star_F=F\circ \star$ is not idempotent. In the continuum the two link fields $A_{\tilde l}$ and $A_{F^{-2}(\tilde l)}$ are infinitesimally close, and monopoles can be thought of as carrying the 
fractional  charge $\theta/(2\pi)$ (two times charge $\theta/(4\pi)$). This is the famous Witten effect \cite{Witten:1979ey}.

\subsection{\texorpdfstring{$PSU(N)$}{PSU(N)} non-abelian gauge theory, discrete \texorpdfstring{$\theta$}{theta}-term and the non-abelian Witten effect}

Recently there was a lot of discussion in the literature about gauging the center of the $SU(N)$ non-abelian gauge theory to obtain an $PSU(N)=SU(N)/\mathbb Z_N$ gauge theory (see, e.g., \cite{Aharony:2013hda}, \cite{Kapustin:2014gua}, \cite{Gaiotto:2017yup}). A prescription to do this is quite straightforward, and was discussed in detail in these papers. Here we mention just the highlights. An $SU(N)$ gauge theory on a cubic lattice is regularized in the usual way with action
\be
S=-\frac{\beta}{2N}\sum_{p}(V_p+c.c.)\;,
\ee
where $V_p$ is the traced plaquette 
\be
V_p=\text{Tr}\prod_{l\in\partial p}U_l\;,
\ee
with $U_l\in SU(N)$, and the trace is over the color index.
The system has a $1$-form center symmetry given by 
\be\label{eq:center_gauge_trans}
U_l\rightarrow e^{i\delta_l}U_l\;,\qquad \text{for }\;\; e^{i\delta_l}\in \mathbb Z_N \; ,
\ee
such that
\be\label{eq:ddelta}
(d\delta)_p=0\bmod \;2\pi\;,\forall p\;.
\ee
This is a $1$-form $\mathbb Z_N$ center symmetry.

Now we can gauge the center symmetry by introducing a $\mathbb Z_N$ valued plaquette field $e^{i B_p}\in \mathbb Z_N$, which is parametrized by a 2-form gauge field $B_p$. The $\mathbb Z_N$ center symmetry is now a gauge symmetry because under the transformation 
\begin{align}
&B_p\rightarrow B_p-(d\delta)_p\\
&U_l\rightarrow U_le^{i\delta_l}
\end{align}
the action is invariant. Notice however that $B_p$ also has a $2$-form gauge symmetry 
\be\label{eq:B_NA_gauge_symm}
B_p\rightarrow B_p+2\pi m_p\;.
\ee 
This gauge symmetry can be thought of as an equivalence of $\delta_l\sim \delta_l+2\pi$ of the gauge transformation parameter $\delta_l$. However, it is more practical to treat $\delta_l$ as a single-valued parametrization of the $\mathbb Z_N$ 1-form gauge transformation, and separately treat \eqref{eq:B_NA_gauge_symm} as a gauge symmetry.

Now we can define a monopole operator of charge $m=1,\dots ,N-1$, as an object for which $(dB)_c$
\be\label{eq:dBN_const}
(dB)_{c}=2\pi m/N \bmod\; 2\pi\;.
\ee
We can impose this constraint by inserting a Lagrange multiplier at every cube $c$
\be\label{eq:dBN_const_LM}
\sum_{a_{\star c}\in \mathbb Z}e^{i a_{\star c}(dB)_{c}}\;.
\ee
The sum over $a_{\star c}$ imposes the constraint that there are no $\mathbb Z_N$ monopoles. We have redundantly summed over all integers, but the $a$-field has a gauge-shift symmetry $a_{\tilde l}\rightarrow a_{\tilde l}+2\pi  Nz_{\tilde l}$, for $z_{\tilde l}\in\mathbb Z$. The field $a$ can therefore be thought of as a (magnetic) $\mathbb Z_N$ gauge field. 

Now if we want to couple (electric) matter to this theory, we must use $PSU(N)$ representations only. This is because the link-fields $U_l$ which are in the representation of $SU(N)$, but not in the representation of $PSU(N)$, are not gauge invariant under the center-gauge symmetry \eqref{eq:center_gauge_trans} and we cannot use them to couple electric matter.

We now define the discrete $\theta$-term as
\be
S_k=\frac{ik N}{4\pi}\sum_{p}B_pB_{F(\star p)}\; .
\ee
We need to choose $k$ such that $B_p\rightarrow B_p+2\pi m_p$ is a gauge symmetry. We have
\be\label{eq:B_gauge_trans}
\Delta S_k=i {k N}\sum_p \left(\frac{m_p B_{F(\star p)}+B_p m_{F(\star p)}}{2}+\pi m_p m_{F(\star p)}\right) \; .
\ee
The above expression is obviously a $2\pi i$ multiple of an integer when $k$ is even. We also must have invariance when we let $B_p\rightarrow B_p-(d\delta)_p$ for $\delta_l=0,2\pi/N,\dots, 2\pi(N-1)/N$. We have that upon partial integration
\be
\Delta S_k=-\frac{i k N}{4\pi} \sum_l \left(\tilde\delta_{\tilde l}+\tilde\delta_{F^{-2}(\tilde l)}\right) (dB)_{\star \tilde l} \; ,
\ee
where $\tilde \delta_{\tilde l}=\delta_{F(\tilde l)}$.
Indeed the above is clearly zero $\bmod\;  2\pi i$ if we impose that there are no monopoles (i.e., $m=0$ in \eqref{eq:dBN_const}), and $k$ is even.

Now consider imposing the constraint  $dB=0\bmod 2\pi$ with the Lagrange multiplier term \eqref{eq:dBN_const_LM}. 
To maintain the 1-form gauge invariance, we must also shift  $a_{l}$ as
\be
a_{\tilde l}\rightarrow a_{\tilde l}+\frac{k N}{4\pi}\left(\delta_{F(\tilde l)}+\delta_{F^{-1}(\tilde l)}\right)\;.
\ee
We could now couple monopole matter, by coupling the lattice $\mathbb Z_N$ gauge field $a_{\tilde l}$ to a, say, scalar field on the dual lattice. If $k=0$ then we can do this with the term
 
\be
(\phi^m_{{\tilde x+\mu}})^*e^{\frac{-2\pi i a_{\tilde x,\mu}}{N}}\phi^m_{\tilde x}+c.c.
\ee
But if $k\ne 0$, the monopole must be a color multiplet. Indeed let $U_l(\mathcal R_s)$ be an $SU(N)$ link field in the representation $\mathcal R_{s}$, where $s$ signifies the number of boxes in the Young tableau of the representation (we refer to such a representation as a $s$-index representation). Under the 1-form gauge symmetry transformation we have that $U_l(\mathcal R_s)\rightarrow e^{is\delta _l} U_l(\mathcal R_s)$. Consider the coupling of magnetic matter in the form
\be
(\phi^m_{\tilde x+\mu})^\dagger U_{F(\tilde l)}(\mathcal R_{s_1})U_{F^{-1}(\tilde l)}(R_{s_2})e^{-\frac{2\pi i a_{\tilde x,\mu}}{N} }\phi^m_{\tilde x}+c.c. \; ,
\ee
where $\phi^m$ is a monopole field containing indices in the $s_1$-index and $s_2$-index representation such that $s_1=k/2\bmod N$ and $s_2=k/2\bmod N$. This is the non-abelian version of the Witten effect discussed previously.

Note that this also gives another insight why in this  lattice regularization we could not have odd $k$. If this was the case we could not couple monopole operators, as we did above, as we need to be able to split the gauge-representation into two parts. Let us comment a bit more about this. The discrete $\theta$-term labeled by $k$ can be thought of as changing a theory where monopoles can only be charged electrically in the $PSU(N)$ representation ($k$=0) to a theory where monopoles can carry a $k$-index representation (modulo $N$). So by setting $k=N$ we return back to the original $k=0$ case where only $PSU(N)$ representations are allowed. However, the construction here requires us to split the electric charge of the monopole to two different links of the lattice. For odd $N$ this is not an issue as even $k=0,2 \dots, 2N$ are enough to construct all the representations from $s_1,s_2=k/2\bmod\; N$ (i.e., if $k=2, 4, 6, N-1$, then we can make a $k$-index representation, while if $k=N+1, \dots, 2N-1$ we can make $1,3, \dots N-2$ representations). However, if $N$ is even, we only have $N/2$ distinct theories, but expect $N$ distinct theories. It is an interesting question whether our construction can be modified to accommodate the missing values of the discrete $\theta$-angle.

\section{Summary and outlook}

In this paper have discussed a particular discretization of abelian gauge theories on the lattice. The construction was motivated by reducing the center symmetry group of the non-compact abelian gauge theory from $\mathbb R$ to $U(1)=\mathbb R/\mathbb Z$, by introducing the $\mathbb Z$-valued 2-form gauge fields living on lattice plaquettes. This construction has many benefits over the conventional $U(1)$ gauge theory construction. Firstly, in addition to the electric matter, one can couple magnetic matter to the theory in a natural way. Secondly, because the action is Gaussian in the gauge degrees of freedom it allows for a natural electric-magnetic duality, even in the presence of matter. This enables one to formulate self-dual abelian gauge theories in four dimensions which is free from sign problems, as long as the matter is bosonic. The construction also allows a natural way to introduce $\theta$-terms in $2d$ and to constrain monopole charges in $3d$ (and also in higher dimensions). These low dimensional examples are important effective theories for anti-ferromagnetic materials. 

We also showed that this logic can be used to construct natural $\theta$ terms in $4d$ theories. The construction parallels that of the work \cite{Kapustin:2014gua} where a discrete $\theta$-term was constructed for $PSU(N)=SU(N)/\mathbb Z_N$ gauge theories on the triangular lattice (i.e., simplicial complexes). We showed here that the construction goes through on cubic lattices as well. While at the moment we did not attempt to solve the complex-action problem in these cases, several avenues exist which may be useful for a numerical study of these theories. One is to notice that the $\theta$-term can be reduced to a single $4d$ hypercube by a gauge transformation. It is  therefore possible that an efficient algorithm can be used which will sum over the degrees of freedom at this spacetime-point, without paying a prohibitively large cost as the infinite volume limit is taken.

Furthermore there remains a question on what can be done in $4d$ $SU(N)$ gauge theories with the conventional (i.e., continuous) $\theta$-term. An interesting avenue to explore here would be to construct an $SU(N)$ lattice gauge theory on a trivial $SU(N)$ bundle. In \cite{Seiberg:2010qd} (see also \cite{Bachas:2016ffl}) such a theory was constructed  by constraining the instantons to be multiples of a particular charge $q$. The theory on a trivial gauge bundle is then  obtained in the limit $q\rightarrow \infty$. But such a theory can also be constructed by viewing the gauge-fields on the links as algebra-valued, not group-valued. Such theories have a $\mathbb Z$ 3-form global symmetry, under which an integral over the Chern-Simons form is charged. The symmetry transformation can be thought of as the would-be large gauge transformation which has a non-trivial winding over a closed 3-cycle, which shifts the Chern-Simons integral by discrete steps. The transformation is however not a gauge transformation on a theory formulated on a trivial $SU(N)$ bundle, but is a global symmetry transformation. The problem then becomes one of finding an expression for a Chern-Simons form on the lattice in terms of the algebra valued fields, such that its lattice exterior derivative gives a topological charge. A construction of this seems to have been made in \cite{Seiberg:1984id} for usual lattice gauge theories with group-valued gauge fields. If this is possible to do for the theory on the trivial gauge bundle, then a similar construction to the one we showed here can be performed for non-abelian gauge theories, where a 4-form discrete gauge field $B_h$ living on hypercubes $h$ is introduced to turn a $\mathbb Z$ 3-form global symmetry to a local one. The $\theta$-term would simply be proportional to the sum $\sum_h B_h$.

\vskip5mm
\noindent
{\bf Acknowledgments:}
\vskip2mm
\noindent
We would like to thank Maria Anosova, Costas Bachas, Zohar Komargodski, Bruno Le Floch, and Yuya Tanizaki 
for the discussions and comments on the draft, especially Bruno Le Floch for his detailed reading 
of the draft. We would also like to thank Daniel G\"oschl for discussions about computational aspects 
of the dual representations. This work is supported by the Austrian Science Fund FWF, grant I 2886-N27. 
The authors would like to express a special thanks to the Mainz Institute for Theoretical Physics (MITP) 
for its hospitality and support during the workshop ''Progress in Diagrammatic Monte Carlo Methods'' 
which facilitated this collaboration.

\vskip15mm

\begin{appendix}

\noindent
In the subsequent appendices we collect several somewhat technical parts needed for the presentation and in 
particular for the transformation of our example systems to worldline and worldsheet representations.

\section{Differential forms on the lattice}\label{app:forms}

In this appendix we collect our conventions for differential forms on a cubic lattice which we use for parts of 
our presentation. For a more general introduction based on triangulated manifolds see, e.g., \cite{wallace}.

We consider a $d$-dimensional cubic lattice $\Lambda$ with lattice constant 1, which is either infinite or has 
periodic boundary conditions. The sites are denoted as $x = (x_1,x_2 \, ... \, x_d)$ with $x_i$ either in all of 
$\mathbb{Z}$ or $x_i = 1, 2 \, ... \, N_i$ with periodic boundary conditions. In addition we periodically identify 
links, plaquettes et cetera. In general we define 
$r$\emph{-cells} $c^{(r)}(x)_{\mu_1\mu_2 \, ... \, \mu_r}$ with $\mu_1 < \mu_2 < \, ... \, < \mu_r$ 
as sets of sites $c^{(r)}(x)_{\mu_1 \mu_2 \, ... \, \mu_r} = 
\{x\} \cup \{x \!+\! \hat{\mu}_i, 1 \leq i \leq r\} \cup \{x \!+\! \hat{\mu}_i \! +\! \hat{\mu}_j, 1 \leq i < j \leq r\} \cup ... 
\cup \{x\!+\!\hat{\mu}_1\! + ... + \!\hat{\mu}_r \}$, where $\hat{\mu}$ denotes the unit vector in direction 
$\mu$. Obviously 0-cells are the sites, 1-cells are the links, 2-cells the plaquettes et cetera. In our basic definition, the 
cells are labelled by the set of canonically ordered directions $\mu_1 < \mu_2 < \, ... \, < \mu_r$, 
but it is convenient to introduce
a more general labelling via $c^{(r)}(x)_{\mu_1 \, ... \, \mu_i \, ... \, \mu_j \, ... \, \mu_r} = - 
c^{(r)}(x)_{\mu_1 \, ... \, \mu_j \, ... \, \mu_i \, ... \, \mu_r}$, where the sign is referred to as the 
\emph{orientation} of the cell.

We define the \emph{boundary operator} $\partial$ acting on $r$-cells $c^{(r)}$ to be the operator that gives the oriented 
sum of the $c^{(r-1)}$ cells in the boundary of $c^{(r)}$, 
\be
\partial c^{(r)}(x)_{\mu_1 \, ... \, \mu_r} = \sum_{k=1}^r (-1)^{k+1} \!
\left[c^{(r-1)}(x\!+\!\hat{\mu}_k)_{\mu_1 \, ... \, {\mu}^{\!\!\!^o}_k \, ... \, \mu_r} \! - 
c^{(r-1)}(x)_{\mu_1 \, ... \, {\mu}^{\!\!\!^o}_k \, ... \, \mu_r} \right] ,
\ee
where ${\mu}^{\!\!\!^o}_k$ indicates that this index is omitted. For 0-cells, i.e., sites, we define 
$\partial \, c^{(0)}(x) = 0$.
We also define the \emph{co-boundary operator} $\hat\partial$ acting on an $r$-cell $c^{(r)}$ to be the operator that  
gives the oriented sum of all $r+1$ cells which contain $c^{(r)}$ in their boundary, 
\be
\hat\partial \, c^{(r)}(x)_{\mu_1 \, ... \, \mu_r}  = \sum_{\nu \neq \mu_1 \, ... \, \mu_r} \left[
c^{(r+1)}(x)_{\mu_1 \, ... \, \mu_r \nu} - c^{(r+1)}(x-\hat{\nu})_{\mu_1 \, ... \, \mu_r \nu} \right],
\ee
with the convention $\hat\partial \, c^{(d)}(x)_{1 \, ... \, d} = 0$. It is easy to show that the boundary and co-boundary 
operators are nilpotent, i.e., $\partial^2 = 0$ and $\hat\partial^2 =0$.

The \emph{dual lattice} $\tilde\Lambda$ is the cubic lattice that has its sites $\tilde{x}$ at the centers of the $d$-cells
of $\Lambda$, i.e., $\tilde{x} = x + \frac{1}{2} ( \hat{1} + \hat{2} + ... + \hat{d})$. There is a natural identification
of the $r$-cells $c^{(r)}$ of $\Lambda$  with the $(d\!-\!r)$-cells $\tilde{c}^{\,(d-r)}$ of $\tilde\Lambda$.
The $\star$ \emph{dual operator} implements this identification, taking into account also the orientation
of the cells,
\begin{subequations}
\begin{multline}
\star \, c^{(r)}(x)_{\mu_1 \, ... \, \mu_r}   =\frac{1}{(d\!-\!r)!} \!\!\!
\sum_{\mu_{r+1}^\prime  ... \, \mu_{d}^\prime} \!\!\!\!\!\!
\epsilon_{\mu_1 \, ... \, \mu_r \mu_{r+1}^\prime \, ... \, \mu_d^\prime} \,
\\\times\tilde{c}^{(d-r)}(\tilde{x} \! - \! \hat{\mu}_{r+1}' ... \! - \! \hat{\mu}_{d}')_{\mu_{r+1}^\prime  ... \, \mu_{d}^\prime} \, ,
\end{multline}
\begin{multline}
\star \, \tilde c^{(r)}(\tilde x)_{\mu_1 \, ... \, \mu_r} = \frac{1}{(d\!-\!r)!} \!\!\!
\sum_{\mu_{r+1}^\prime  ... \, \mu_{d}^\prime} \!\!\!\!\!\!
\epsilon_{\mu_1 \, ... \, \mu_r \mu_{r+1}^\prime \, ... \, \mu_d^\prime} \,
\\\times c^{(d-r)}(x \! + \! \hat{\mu}_{1} ... \!+ \! \hat{\mu}_{r})_{\mu_{r+1}^\prime  ... \, \mu_{d}^\prime} .
\end{multline}
\end{subequations}
It is easy to show that up to a sign (only for even dimensions) 
the square of the $\star$ operator is the identity, $\star^2 \, c^{(r)} \, = \; (-1)^{r(d-r)} c^{(r)}$, 
$\star^2 \, \tilde c^{(r)} \, = \; (-1)^{r(d-r)} \tilde c^{(r)}$.

Also notice that we can write the co-boundary operator as
\be
\hat\partial\, c^{(r)} \; \, = \; \, (-1)^{(d-r-1)(r+1)}  \star \partial \star \, c^{(r)} \; .
\ee
or, equivalently
\be
\partial\star=\star \hat\partial
\ee

We define an $r$\emph{-form} $A_{c^{(r)}}$ (often also referred to as $r$\emph{-chain}), 
to be an object that lives on an $r$-cell.  
When we refer to a specific $r$-cell as $c^{(r)}(x)_{\mu_1 \mu_2 \, ... \, \mu_r}$, with 
$\mu_1 <  \mu_2 < ... < \mu_r$, we also use the notation
\begin{equation}
A(x)_{\mu_1 \, ... \, \mu_r}  \quad \mbox{with}  \quad 
A(x)_{\mu_1 \, ... \, \mu_i \, ... \, \mu_j \, ... \, \mu_r} \; = \; - A(x)_{\mu_1 \, ... \, \mu_j \, ... \, \mu_i \, ... \, \mu_r}
\; ,
\label{r_form}
\end{equation}
to denote the $r$-forms and in the second part of (\ref{r_form}) already have introduced our convention
for the orientation of $r$-forms, i.e., $A_{-c^{(r)}} = - A_{c^{(r)}}$.  

The star operator also induces a map from $r$-forms on $\Lambda$ to $(d-r)$-forms on $\tilde \Lambda$. For an $r$-form on $\Lambda$ we define an $(d-r)$-form $\star A$ on the dual lattice as
\be
(\star A)_{\tilde c^{(d-r)}}= A_{\star \tilde c^{(d-r)}}
\ee

Also for the action of the $\star$ dual operator on 
$r$-forms one finds the relation $\star^2 A_{c^{(r)}} = (-1)^{r(d-r)} A_{c^{(r)}}$, 
$\star^2 \tilde A_{\tilde c^{(r)}} = (-1)^{r(d-r)} \tilde A_{\tilde c^{(r)}}$.

The \emph{exterior derivative} operator $d$ maps an $r-1$ form $A_{c^{(r-1)}}$ to an $r$ form $(d A)_{c^{(r)}}$
(with the convention $d A_{c^{(d)}} = 0$),
\begin{eqnarray}
(d A)_{c^{(r)}} & \!\! = \!\! & \!\!\! \sum_{c^{(r-1)} \in \partial c^{(r)}} \!\!\!  A_{c^{(r-1)}} \; ,
\\
(d A)(x)_{\mu_1 \, ... \, \mu_{r}} & \!\! = \!\! & \sum_{k=1}^{r} (-1)^{k+1} \!
\left[A(x\!+\!\hat{\mu}_k)_{\mu_1 \, ... \, {\mu}^{\!\!\!^o}_k \, ... \, \mu_{r}} \! - 
A(x)_{\mu_1 \, ... \, {\mu}^{\!\!\!^o}_k \, ... \, \mu_{r}} \right] \; .
\nonumber 
\end{eqnarray}
It is straightforward to show that the exterior derivative operator is nilpotent, i.e., $d^{\,2} = 0$.

The \emph{boundary operator} $\delta$ for $r$-forms (often also referred to as \emph{divergence operator}) maps
an $r+1$-form $A_{c^{(r+1)}}$ to an $r$ form $(\delta A)_{c^{(r)}}$ 
(with the convention $\delta A_{c^{(0)}} = 0$),
\begin{eqnarray}
(\delta A)_{c^{(r)}} & \!\! = \!\! & \!\!\! \sum_{c^{(r+1)} \in \hat \partial c^{(r)}} \!\!\!  A_{c^{(r+1)}} \; ,
\\
(\delta A)(x)_{\mu_1 \, ... \, \mu_{r}} & \!\! = \!\! & \sum_{\nu} 
\left[ A(x)_{\mu_1 \, ... \, \mu_r \, \nu}  - A(x-\hat\nu)_{\mu_1 \, ... \, \mu_r \, \nu}  \right] \; .
\nonumber 
\end{eqnarray}
Also here nilpotency is straightforward to show, i.e., $\delta^2 = 0$. The operator $\delta$ can also be expressed as $\delta c^{(r)}=(-1)^{(d-r+1)(r+1)}\star d\star c^{(r)}$ or $\star\delta=d\star$. 

The exterior derivative operator $d$ and the boundary operator $\delta$ are dual to each other with respect to 
the $l_2$ norm. In other words, they obey the \emph{partial integration formula}
\begin{equation}\label{eq:partial_int}
\sum_{c^{(r)}} \,  (d A)_{c^{(r)}} \, B_{c^{(r)}} \; = \; (-1)^{r} \sum_{c^{(r-1)}} \,  A_{c^{(r-1)}} \, 
(\delta B)_{c^{(r-1)}} \; .
\end{equation}

Another form we will use involves a product of forms on the lattice and the dual lattice. We have 
\be
\sum_{c^{(r)}} (dA)_{c^{(r)}}\, \tilde B_{\star c^{(r)}}\;= \; (-1)^{r} \sum_{c^{(r-1)}} \,  A_{c^{(r-1)}} \, 
( d\tilde B)_{\star c^{(r-1)}} \; .
\ee
To derive the above formula we used $d\star=\star \delta$ in  \eqref{eq:partial_int}, with $\tilde B_{\tilde c^{(d-r)}}=(\star B)_{\tilde c^{(d-r)}}$.

We finish by quoting the \emph{Hodge decomposition} formula. For any $r$-form $A_{c^{(r)}}$ there exist an $r+1$ form $B_{c^{(r+1)}}$,
an $r-1$ form $C_{c^{(r-1)}}$ and a harmonic (defect) $r$-form $H_{c^{(r)}}$, which obeys
$d \, H_{c^{(r)}} = \delta \, H_{c^{(r)}} = 0$, such that 
\begin{equation}
A_{c^{(r)}} \; = \; \delta \, B_{c^{(r+1)}} \; + \; d \, C_{c^{(r-1)}} \; + \; H_{c^{(r)}} \; .
\end{equation}

\section{Poisson summation formula}

For completeness of our presentation and self-consistency of the paper we here prove the well-known result
\be
b(F) \; \equiv \;  \sum_{n \in \mathbb{Z}} 
e^{ -  \frac{\beta}{2}\left(F+2\pi n\right)^2  -  i \gamma \left(F+2\pi n\right)}  =\frac{1}{\sqrt{2\pi \beta}} \; \sum_{p \in \mathbb{Z}} \; e^{i p F} e^{- \frac{1}{2\beta} \left(p + \gamma \right)^2} \; .
\ee
All gauge field Boltzmann factors we consider in this paper are products of 
terms $b(F)$ that depend on a single component $F \equiv F_{x,\mu\nu}$ of the field 
strength tensor at a fixed site $x$, where $\gamma$ is some real number.
Obviously the function $b(F)$ is $2\pi$-periodic in $F$ such that it has the Fourier representation 
\begin{equation}
b(F) \; = \; \sum_{p \in \mathbb{Z}}  \widetilde{b}(p) \; e^{i p F} \; .
\label{Frep}
\end{equation}
The Fourier transform $\widetilde{b}(p)$ can be evaluated in closed form,
\begin{eqnarray}
\widetilde{b}(p) & \!=\! & \int_{-\pi}^\pi \! \frac{dF}{2\pi} \; b(F) \; e^{-i p F} \; = \; 
\sum_{n \in \mathbb{Z}}  \int_{-\pi}^\pi \! \frac{dF}{2\pi}  \; 
e^{ -  \frac{\beta}{2}\left(F+2\pi n\right)^2  -  i \gamma \left(F+2\pi n\right)} \; e^{-i p F} 
\nonumber \\
& \!=\! & 
\sum_{n \in \mathbb{Z}}  \int_{-\pi}^\pi \! \frac{dF}{2\pi} \; 
e^{ -  \frac{\beta}{2}\left(F+2\pi n\right)^2  -  i \gamma \left(F+2\pi n\right) -  i p \left(F+2\pi n\right)} 
\; = \; \Big| F^\prime \!= F\!+\!2\pi n \Big|
\nonumber \\
& \!=\! & 
\sum_{n \in \mathbb{Z}}  \int_{2\pi n -\pi}^{2\pi n + \pi} \! \frac{dF^\prime}{2\pi} \; 
e^{ -  \frac{\beta}{2}\left(F^\prime\right)^2  -  i F^\prime\left(p + \gamma \right)} 
\nonumber \\
& \!=\! & 
\int_{-\infty}^{\infty} \! \frac{dF^\prime}{2\pi} \; 
e^{ -  \frac{\beta}{2}\left(F^\prime\right)^2  -  i F^\prime\left(p + \gamma\right)}
\; = \; \frac{1}{\sqrt{2\pi \beta}} \; e^{- \frac{1}{2\beta} \left(p + \gamma \right)^2} \; .
\end{eqnarray}
Inserting this Fourier transform $\widetilde{b}(p)$ in (\ref{Frep}) gives rise to the Fourier representation 
of the Boltzmann factor used for the mapping to dual representations.

\section{Worldline representation for the \texorpdfstring{$\phi^4$}{phi-4} field in a gauge background}\label{app:Higgs}

In this section we briefly outline the steps for obtaining the worldline representation for the 
partition sum $Z_e[U]$ of the $\phi^4$ field in a background gauge configuration $U$.
For a detailed derivation we refer to \cite{Mercado:2013yta}.

In terms of conventional fields the partition sum $Z_e[U]$ in a background of the group-valued link 
variables is defined in (\ref{ZM_lattice}). Rewriting the sums in the action (\ref{S_Higgs}) that appears in the 
exponent of the Boltzmann factor we find (here we set $V_e( |\phi_x| ) = \lambda |\phi_x|^4$ and drop 
some of the subscripts)
\begin{eqnarray}
Z_e^{\, \lat}[U] & \!\!\!\!\! =  \!\!\!\!\! & \int \!\! D[\phi] \prod_x e^{- M |\phi_x|^2 - \lambda |\phi_x|^4}
\prod_{x,\mu} e^{\, \phi_x^* U_{x,\mu} \phi_{x+\hat\mu}}
e^{\, \phi_x U_{x,\mu}^{\, *} \phi_{x+\hat\mu}^*}
\\
& \!\! \!\!\!  =  \!\!\!\!\! & \!\!\sum_{\{ n, \overline{n} \}}  \!
\int \!\! D[\phi] \prod_x \! e^{- M |\phi_x|^2 - \lambda |\phi_x|^4} \!
\prod_{x,\mu}   \!
\frac{ \!\big( \phi_x^* U_{x,\mu} \phi_{x+\hat\mu} \big)^{\!n_{x,\mu}} \!
\big( \phi_x U_{x,\mu}^* \phi_{x+\hat\mu}^* \big)^{\!\overline{n}_{x,\mu}}}
{n_{x,\mu}! \; \; \overline{n}_{x,\mu} !} .
\nonumber 
\end{eqnarray} 
In the second line we have expanded the exponentials for the nearest neighbor terms 
and introduced the notation $\sum_{\{ n, \overline{n} \}} = 
\prod_{x,\mu} \sum_{n_{x,\mu} \in \mathbb{N}_0} 
\sum_{\overline{n}_{x,\mu} \in \mathbb{N}_0}$ for the multi-sum over the corresponding summation 
indices. 

Next we organize the powers of the fields $\phi_x$ and $\phi_x^\star$ with respect to sites
and subsequently switch to polar coordinates $\phi_x = r_x e^{i \alpha_x}$. The partition 
sum turns into (we define $f_x \equiv \sum_\mu [n_x + \overline{n}_x
+ n_{x-\hat{\mu}} + \overline{n}_{x-\hat{\mu}}]$)
\begin{eqnarray}
Z_e^{\, \lat} [U] & \!\!   =  \!\! & \!\!\sum_{\{ n, \overline{n} \}}  \prod_{x,\mu}
\frac{ \!\big( U_{x,\mu} \big)^{\!n_{x,\mu} - \, \overline{n}_{x,\mu}}}
{n_{x,\mu}! \; \; \overline{n}_{x,\mu} !} \; \prod_x \int_0^{\infty} \!\!\!\! dr_x \, r_x^{\; f_x + 1} 
\, e^{ \, - M r_x^2 \, - \, \lambda r_x^4} 
\nonumber \\
&& \times \prod_x \int_0^{2\pi}  \frac{d \alpha_x}{2\pi} \; e^{ \, i \alpha_x \,  
\sum_\mu [n_x \, + \, \overline{n}_x \, - \, n_{x-\hat{\mu}} \, - \, \overline{n}_{x-\hat{\mu}}]} \; ,
\end{eqnarray}
where we use the measure in polar coordinates $\int \! D[U] = \prod_x \!  \int_0^{2\pi} \! \frac{d \alpha_x}{2\pi}
\int_0^{\infty} \! dr_x \, r_x$. The integral over the moduli $r_x$ obviously gives rise to the integral weight $I(f_x)$ defined 
in (\ref{W_e_$2d$GH}) and (\ref{W_e_$3d$GH}), while the integral over the phases $\alpha_x$ generates 
the constraints 
$\sum_\mu [n_{x,\mu} \, + \, \overline{n}_{x,\mu} \, - \, n_{x-\hat{\mu},\mu} \, - \, \overline{n}_{x-\hat{\mu},\mu}]  = 
0 \; \forall x$. 

The final step is to reorganize the summation variables by introducing the flux variables $k_{x,\mu} \in 
\mathbb{Z}$ and the auxiliary variables $a_{x,\mu} \in \mathbb{N}_0$, such that 
\begin{equation}
n_{x,\mu} \, - \, \overline{n}_{x,\mu} \; = \; k_{x,\mu} \qquad \mbox{and} \qquad 
n_{x,\mu} \, + \, \overline{n}_{x,\mu} \; = \; | k_{x,\mu} | + 2 a_{x,\mu} \; .
\end{equation}
The multiple sum $\sum_{\{ n, \overline{n} \}}$ is replaced by two sums 
$\sum_{ \{ k\}} = \prod_{x,\mu} \sum_{k_{x,\mu} \in \mathbb{Z}}$ and 
$\sum_{ \{ a\}} = \prod_{x,\mu} \sum_{a_{x,\mu} \in \mathbb{N}_0}$. The constraints can be written as a product over 
Kronecker deltas which we here denote as $\delta(n) \equiv \delta_{n,0}$. They enforce the zero divergence
condition  $\vec{\nabla}\cdot \vec{k}_x = \sum_{\mu} [ k_{x,\mu} - k_{x - \hat{\mu},\mu} ] = 0$ $\forall x$
for the flux variables $k_{x,\mu}$. The partition sum thus can be written in the worldline form
\begin{equation}
Z_e^{\, \lat} [U] \; = \; \sum_{\{k\}} W_e[k] \, \prod_x \delta\left(\vec{\nabla}\cdot \vec{k}_x\right) 
\, \prod_{x,\mu} \, \big( U_{x,\mu} \big)^{\,k_{x,\mu}} \; ,
\end{equation}
where all weight factors and the sum $\sum_{ \{ a\}}$ over the auxiliary variables were collected in the 
real and positive weight $W_e[k]$, which is given in its explicit form in (\ref{W_e_$2d$GH}) and (\ref{W_e_$3d$GH}).

\section{Worldline representation for the \texorpdfstring{$CP(N-1)$}{CP-(N-1)} field}\label{app:CPN}

In this appendix we provide a brief summary of the steps that lead to the worldline representation of the partition
function $Z_e[U]$ in a background configuration of the compact link variables $U_{x,\mu}$ 
(see \cite{Bruckmann:2015sua} for a complete derivation). The conventional partition function $Z_e[U]$ is given by a straightforward discretization of the continuum action \eqref{eq:CPN_action}, and again the first step is to write all sums that appear in the action, i.e., the exponent 
of the Boltzmann factor as a product over the individual Boltzmann factors,
\begin{eqnarray}
Z_e[U] & \!\! = \!\! & \int \! D[ \vec{z} \, ] \; \prod_{x,\mu} \prod_{j = 1}^N \, e^{\, J z_{x,j}^* \, U_{x,\mu} \, z_{x+\hat \mu,j}}
\, e^{\, J z_{x,j} \, U_{x,\mu}^{\, *} \, z_{x+\hat \mu,j}^{\, *}} 
\nonumber \\
& \!\! = \!\! &  \sum_{\{n, \overline{n}\}} \prod_{x,\mu} 
\big( U_{x,\mu} \big)^{ \, \sum_{j=1}^N [ n_{x,\mu}^{(j)} - \, \overline{n}_{x,\mu}^{(j)}] } \; \; 
\prod_{x,\mu} \prod_{j=1}^N
\frac{  J^{\,n_{x,\mu}^{(j)} \, +\, \overline{n}_{x,\mu}^{(j)}}}{n_{x,\mu}^{\,(j)} \, ! \; \; \overline{n}_{x,\mu}^{\, (j)} \, !} 
\nonumber \\
& & \times
\int \! D[ \vec{z} \, ] \; \prod_{x,\mu} \prod_{j = 1}^N 
\big( z_{x,j}^* \, z_{x+\hat \mu,j} \big)^{\, n_{x,\mu}^{\,(j)}} \, 
\big( z_{x,j} \, z_{x+\hat \mu,j}^* \big)^{\, \overline{n}_{x,\mu}^{\,(j)}} \; ,
\end{eqnarray}
where in the second step we have again expanded the individual Boltzmann factors. Since here we 
have vectors $\vec{z}_x$ with $N$ components $z_{x,j}$ we need expansion indices 
$n_{x,\mu}^{\,(j)} \in \mathbb{N}_0$ and $\overline{n}_{x,\mu}^{\,(j)} \in \mathbb{N}_0$ for the two 
nearest neighbor terms for each component $j = 1,2, \, ... \, N$. The corresponding multiple sum is denoted 
by $\sum_{ \{ n, \overline{n} \} } = 
\prod_{x,\mu} \prod_{j=1}^N \sum_{n_{x,\mu}^{\,(j)} \in \mathbb{N}_0} 
\sum_{ \overline{n}_{x,\mu}^{\,(j)} \in \mathbb{N}_0}$.

In order to solve the integral over the moments of spin components $z_{x,j}$ one switches to polar coordinates
similar to the case of the $\phi^4$ theory discussed in Appendix B, and sets $z_{x,j} = e^{i \alpha_{x,j}} \, r_{x,j}$,
where the moduli $r_{x,j}$ obey the constraint $\sum_{j = 1}^N r_{x,j}^{\,2} = 1 \; \forall x$, which ensures the 
normalization condition $\sum_{j = 1}^N | z_{x,j} |^2  = 1  \; \forall x$. Thus the $r_{x,j}$ can be parameterized by 
$N-1$ polar angles $\theta_{x,j}, j = 2,3, \, ... \, N$ (see \cite{Bruckmann:2015sua} for an explicit parameterization).

As for the $\phi^4$ case discussed in Appendix B, the integrals over the phases $\alpha_{x,j}$ give rise to 
constraints, which here appear for each component:  
$\sum_\mu [n_{x,\mu}^{\,(j)} \, + \, \overline{n}_{x,\mu}^{\,(j)} \, - \, n_{x-\hat{\mu},\mu}^{\,(j)} \, - \, 
\overline{n}_{x-\hat{\mu},\mu}^{\,(j)}]  = 0\; 
\forall x \; \forall j$. The integrals over the polar angles $\theta_{x,j}$ give rise to real and positive 
combinatorial factors \cite{Bruckmann:2015sua}. 

In a final step we again switch to flux variables $k_{x,\mu}^{\,(j)} \in \mathbb{Z}$ and auxiliary variables
$a_{x,\mu}^{\,(j)} \in \mathbb{N}_0$, which are related to the $n_{x,\mu}^{\,(j)}, \overline{n}_{x,\mu}^{\,(j)}$ via
\begin{equation}
n_{x,\mu}^{\,(j)} \, - \, \overline{n}_{x,\mu}^{\,(j)} \; = \; k_{x,\mu}^{\,(j)} \qquad \mbox{and} \qquad 
n_{x,\mu}^{\,(j)} \, + \, \overline{n}_{x,\mu}^{\,(j)} \; = \; | k_{x,\mu}^{\,(j)} | + 2 a_{x,\mu}^{\,(j)} \; .
\end{equation}
The constraints then take the form of zero divergence conditions for each flux,
$\vec{\nabla}\cdot \vec{k}_x^{\,(j)} = \sum_{\mu} [ k_{x,\mu}^{\,(j)} - k_{x - \hat{\mu},\mu}^{\,(j)}] = 0$ $\forall x$
$\forall j$, and we can write the worldline representation for $Z_e[U]$ in the form
\begin{equation}
Z_e[U] \; = \; \sum_{\{ k \} } W_e[k] \, \prod_{x} \, \prod_{j = 1}^N 
\delta\big( (\vec{\nabla} \cdot \vec{k}^{\,(j)})_x \big) \, 
\prod_{x,\mu} \big( U_{x,\mu} \big) ^{\, \sum_{j=1}^N k_{x,\mu}^{(j)} } \; .
\label{cpn_expansioncp2}
\end{equation}
All weight factors were collected in the real and positive weights $W_e[k]$, which are a sum over the configurations
of the auxiliary variables  $a_{x,\mu}^{\,(j)}$. The explicit form of the $W_e[k]$ is given in (\ref{cpn_WM}).

\end{appendix}

\vskip10mm
\bibliographystyle{utphys}
\bibliography{bibliography}

\end{document}